\documentclass[opre,nonblindrev]{informs3_hide}
\DoubleSpacedXI 

\usepackage{bbm}
\usepackage[normalem]{ulem}
\usepackage{hyperref}
\usepackage{cleveref}
\crefname{subsection}{subsection}{subsections}

\usepackage[dvipsnames]{xcolor}
\newcommand{\na}[1]{}

\newcommand{\eps}{\varepsilon}
\newcommand{\bI}{\mathbbm{1}}
\DeclareMathOperator*{\bE}{\mathbb{E}}

\newcommand{\bR}{\mathbb{R}}

\renewcommand{\Pr}{\mathbb{P}}

\newcommand{\Bin}{\mathrm{Bin}}
\newcommand{\Pois}{\mathrm{Pois}}
\newcommand{\vp}{\textbf{p}}

\newcommand{\bV}{{\bf V}}
\newcommand{\Dp}{D_{\vp}}
\newcommand{\bF}{{\bf F}}

\newcommand{\OMN}{\mathcal{PHT}}
\newcommand{\ST}{\mathcal{ST}}
\newcommand{\LP}{\mathcal{LP}}
\newcommand{\ED}{\mathcal{T\hspace{-1.5 pt}D}}

\newcommand{\AR}{\mathsf{AR}}
\newcommand{\MR}{\mathsf{UT}}
\newcommand{\ID}{\mathsf{ID}}

\usepackage{xparse}
\NewDocumentEnvironment{myproof}{o}
  {\IfNoValueTF{#1}{\paragraph{{\normalsize \textit{Proof.}}}} {\paragraph{{\normalsize #1.} }} }
  {\hfill$\square$}

\usepackage{natbib,bbm,multirow,multicol}
\bibpunct[, ]{(}{)}{,}{a}{}{,}%
\def\bibfont{\small}%
\TheoremsNumberedThrough     
\ECRepeatTheorems

\EquationsNumberedThrough    

\MANUSCRIPTNO{OPRE-2021-08-497.R1} 

\begin{document}




\TITLE{Tight Guarantees for Static Threshold Policies in the Prophet Secretary Problem}

\ARTICLEAUTHORS{
\AUTHOR{Nick Arnosti}
\AFF{Department of Industrial and Systems Engineering, University of Minnesota, Minneapolis, MN, \EMAIL{arnosti@umn.edu}}
\AUTHOR{Will Ma}
\AFF{Graduate School of Business, Columbia University, New York, NY 10027, \EMAIL{wm2428@gsb.columbia.edu}}
}

\ABSTRACT{

In the \textit{prophet secretary} problem, $n$ values are drawn independently from known distributions, and presented in a uniformly random order. A decision-maker must accept or reject each value when it is presented, and may accept at most $k$ values in total. The objective is to maximize the expected sum of accepted values.

We analyze the performance of {\em static threshold policies}, which accept the first $k$ values exceeding a fixed threshold (or all such values, if fewer than $k$ exist). We show that an appropriate threshold guarantees $\gamma_k = 1 - e^{-k}k^k/k!$ times the value of the offline optimal solution. Note that $\gamma_1 = 1-1/e$, and by Stirling's approximation $\gamma_k \approx 1-1/\sqrt{2 \pi k}$.  This represents the best-known guarantee for the prophet secretary problem for all $k>1$, and is \textit{tight} for all $k$ for the class of static threshold policies.

We provide two simple methods for setting the  threshold. Our first method sets a threshold such that $k \cdot \gamma_k$ values are accepted in expectation, and offers an optimal guarantee for all $k$. Our second sets a threshold such that the expected number of values exceeding the threshold is equal to $k$. This approach gives an optimal guarantee if $k > 4$, but gives sub-optimal guarantees for $k \le 4$. Our proofs use a new result for optimizing sums of independent Bernoulli random variables, which extends a classical result of \citet{hoeffding_1956} and is likely to be of independent interest. Finally, we note that our methods for setting thresholds can be implemented under limited information about agents' values.

}


\HISTORY{This version from June 7th, 2022.  Full version of EC2022 paper.}

\maketitle


\section{Introduction}\label{sec:intro}

A decision-maker is endowed with $k$ identical, indivisible items to allocate among $n > k$ applicants who arrive sequentially. Each applicant $i=1,\ldots,n$ has a value $V_i\ge0$ for receiving an item. Upon the arrival of applicant $i$, the decision-maker observes $V_i$ and must either immediately ``accept'' and allocate an item to applicant $i$, or irrevocably ``reject'' applicant $i$ (which is the only option if no items remain). The policymaker wishes to maximize the sum of the values of accepted applicants.


\textbf{The Prophet Secretary problem.} This style of sequential accept/reject problem has been studied under different assumptions about the decision-maker's initial information and the order in which applicants arrive. In the {\em secretary problem}, the decision-maker knows nothing about applicants' values, but applicants arrive in a uniformly random order.
Meanwhile, papers on \textit{prophet inequalities} typically assume that each value $V_i$ is drawn independently from a known distribution $F_i$. 
This paper studies the \textit{prophet secretary problem} introduced by \citet{esfandiari2017prophet}, in which values are drawn independently from known distributions and applicants arrive in a uniformly random order.  We  discuss the background behind these naming conventions in Section \ref{sec:history}.

\textbf{Static Threshold Policies.} We focus especially on {\em static threshold policies}, which fix a real number $t$, and accept the first $k$ applicants whose value exceeds $t$. These policies are simple to explain and implement, and are non-discriminatory, in that they use the same threshold for every applicant (so long as items remain). Perhaps for these reasons, static threshold policies arise frequently in practice. They are typically implemented by combining eligibility criteria with first-come-first-served allocation. For example, policymakers often use an income ceiling to determine eligibility for subsidized housing, and then award this housing using a first-come-first-served waitlist. Similarly, many popular marathons set ``qualifying times'', and allow qualified runners to claim race slots until all slots have been taken. 

Static threshold policies have also been widely deployed is the allocation of COVID-19 vaccines. In early 2021, many state 
governments sorted people into priority tiers. At any given time, only the highest-priority tiers were eligible for vaccination. However, eligible individuals could claim appointments on a first-come-first-served basis. 
When determining the eligibility threshold, policymakers balanced two risks. Using very strict criteria could result in unclaimed appointments and discarded doses. Expanding eligibility could result in appointments being claimed as soon as they became available, causing some high-priority individuals to be turned away.\footnote{Both of these concerns arose in the state of New York, which began with very strict criteria that resulted in wasted doses, and then expanded eligibility, causing a scramble for appointments \citep{nytimes-1, nytimes-2}.}


\textbf{Key questions and contributions.} Experiences like this one inspire two natural questions. First, what is the cost of using eligibility criteria alongside first-come-first-served allocation, rather than waiting for all applicants to arrive and then selecting those with the highest priority? 
Second, how should the eligibility threshold be set to balance the two risks described above? We use the prophet secretary model to investigate both questions.

Our first contribution is to exactly characterize the worst-case performance of static threshold policies. More specifically, we show that there always exists a static threshold policy whose performance is at least a $(1 - e^{-k}\frac{k^k}{k!})$-fraction of the full-information benchmark. We also provide an example showing that no better guarantee is possible.

Our second contribution is to show that simple and intuitive algorithms for setting the eligibility threshold achieve optimal guarantees. Our first algorithm sets a threshold based on the expected number of allocated items, while the second is based on the expected number of eligible applicants. In Section \ref{sec:conc}, we note that these algorithms can also be deployed in settings where values are only imperfectly observed, and explain why this is relevant for the allocation of COVID-19 vaccines. 

Despite their simplicity, our algorithms not only provide optimal guarantees within the class of static threshold policies, but also provide the best-known guarantees for {\em any} online policy that observes one applicant's value at a time and must make an irrevocable decision for each applicant before moving on to the next.


\textbf{Terminology and notation.}
Before presenting our results, we define a few concepts.
An \textit{instance} of the prophet secretary problem is given by the number of items $k$ and the number of applicants $n$, along with $n$ cumulative distribution functions $\bF=(F_1,\ldots,F_n)$ on the non-negative real numbers.  
The values $V_i$ are drawn independently from distributions $F_i$, and arrive in a uniformly random order. If $F_i = F_j$ for all $i,j$, we say that values are independently and identically distributed (IID). 

Given values $\{V_i\}_{i = 1}^n$, {\em demand} at a threshold $t$ is the number of values that exceed $t$. The number of accepted applicants is equal to the minimum of demand and the supply $k$. {\em Utilization} is the fraction of the supply that is allocated, and is calculated by dividing the number of accepted applicants by $k$. When using a static threshold policy with threshold $t$, the {\em performance} of $t$ is the expected sum of accepted applicants' values. We compare this performance to the
expected sum of the $k$ highest values, which we call the \textit{prophet's value}. More formal definitions for all of these concepts are provided in Definition \ref{def:value} in Section \ref{sec:a-m-proof}.  Finally, for $k\ge1$, define the constant
\begin{align}
\gamma_k:=1-e^{-k}\frac{k^k}{k!}. \label{eq:gamma}
\end{align}


\subsection{An Upper Bound for Static Threshold Policies} \label{sec:ub}

We first derive an upper bound on the performance for any static threshold policy.

\begin{theorem} \label{thm:upper-bound}
For any $k\ge1$ and $\epsilon > 0$, there exists an instance $(k,\bF)$ with IID values such that the performance of any threshold is less than $\gamma_k+\epsilon$ times the prophet's value.
\end{theorem}

This upper bound is for static threshold policies, and no longer holds if more general policies are allowed. For example, when $k = 1$ we have $\gamma_1 =1-1/e \approx 0.632$, but there exist policies that achieve a guarantee of 0.669 \citep{correa-saona-ziliotto_2020}, and 0.745 if values are IID \citep{correa_foncea_hoeksma_oosterwijk_vredeveld_2017}.

Our proof of Theorem \ref{thm:upper-bound} consists of a careful analysis of the following example.

\begin{example} \label{eg:countereg}
Let the values $V_i$ be independent and identically distributed according to
\begin{align*}
\SingleSpacedXI
V_i=
\begin{cases}
n W_k & w.p.~1/n^2\\
1 & w.p.~1-1/n^2,\\
\end{cases}
\DoubleSpacedXI
\end{align*}
where $W_k = k \frac{\mathbb{P}(\Pois(k) < k)}{\mathbb{P}(\Pois(k) > k)}$, and $\Pois(k)$ denotes a Poisson random variable with mean $k$.
\end{example}
On this example, the prophet always accepts $k$ values, and takes high values whenever they occur. The prophet's value is close to $k+W_k$. When setting a static threshold, the key question is whether to accept applicants with $V_i = 1$. Always accepting these applicants results in performance close to $k$, while rejecting them results in performance close to $W_k$.
We show that even accepting these applicants with some probability $p \in (0,1)$ cannot perform better than $\gamma_k (k + W_k)$.

\subsection{A Matching Lower Bound}

\Cref{eg:countereg} gives an upper bound for the performance of static threshold policies.
We now show that this bound can be attained by setting the threshold so that expected utilization equals $\gamma_k$.


\begin{theorem} \label{thm:a-m}
For any instance $(k,\bF)$, the performance of any threshold such that expected utilization is equal to $k$ is at least $\gamma_k$ times the prophet's value.
\end{theorem}

\begin{remark} 
If each distribution $F_i$ is continuous, then expected utilization decreases continuously from 1 (at $t = 0$) to $0$ (as $t \rightarrow \infty$), so it is always possible to choose a threshold such that expected utilization equals $\gamma_k$. In the case of discontinuous (discrete) distributions, tie-breaking may be required. In this case, we consider policies  defined by a threshold $t \ge 0$ and a tie-break probability $p \in [0,1]$. If $V_i$ is exactly equal to $t$ and items remain when $i$ is considered, then the policy $(t,p)$ accepts item $i$ with probability $p$ (independently from all other randomness). 
Tie-breaking makes it possible to adjust expected utilization (and expected demand) continuously, and is a standard way to handle discrete distributions. For example, it is also used in \citet{ehsani2018prophet} and \citet{chawla-devanur-lykouris_2020}. All of our results apply to any static threshold policy with tie-breaking.
\end{remark}

For any $k>1$, \Cref{thm:a-m} implies the best-known guarantees (relative to the prophet's value) for \textit{any} online policy in the prophet secretary problem. In fact, our proof establishes that the statement remains true if the prophet's value is replaced by a stronger benchmark (sometimes called the ``LP relaxation", ``fluid limit", or ``ex ante relaxation") that is constrained to accept at most $k$ values {\em in expectation}, rather than on each realization. We formally define this benchmark in \Cref{sec:a-m-proof}. It is well-known that no online policy can guarantee performance better than $\gamma_k$ times the value of the LP relaxation \citep[see e.g.][]{yan2011mechanism}, so Theorem \ref{thm:a-m} also implies that static threshold policies achieve optimal guarantees against this benchmark.

\begin{figure}
\centering
\includegraphics[width = \textwidth]{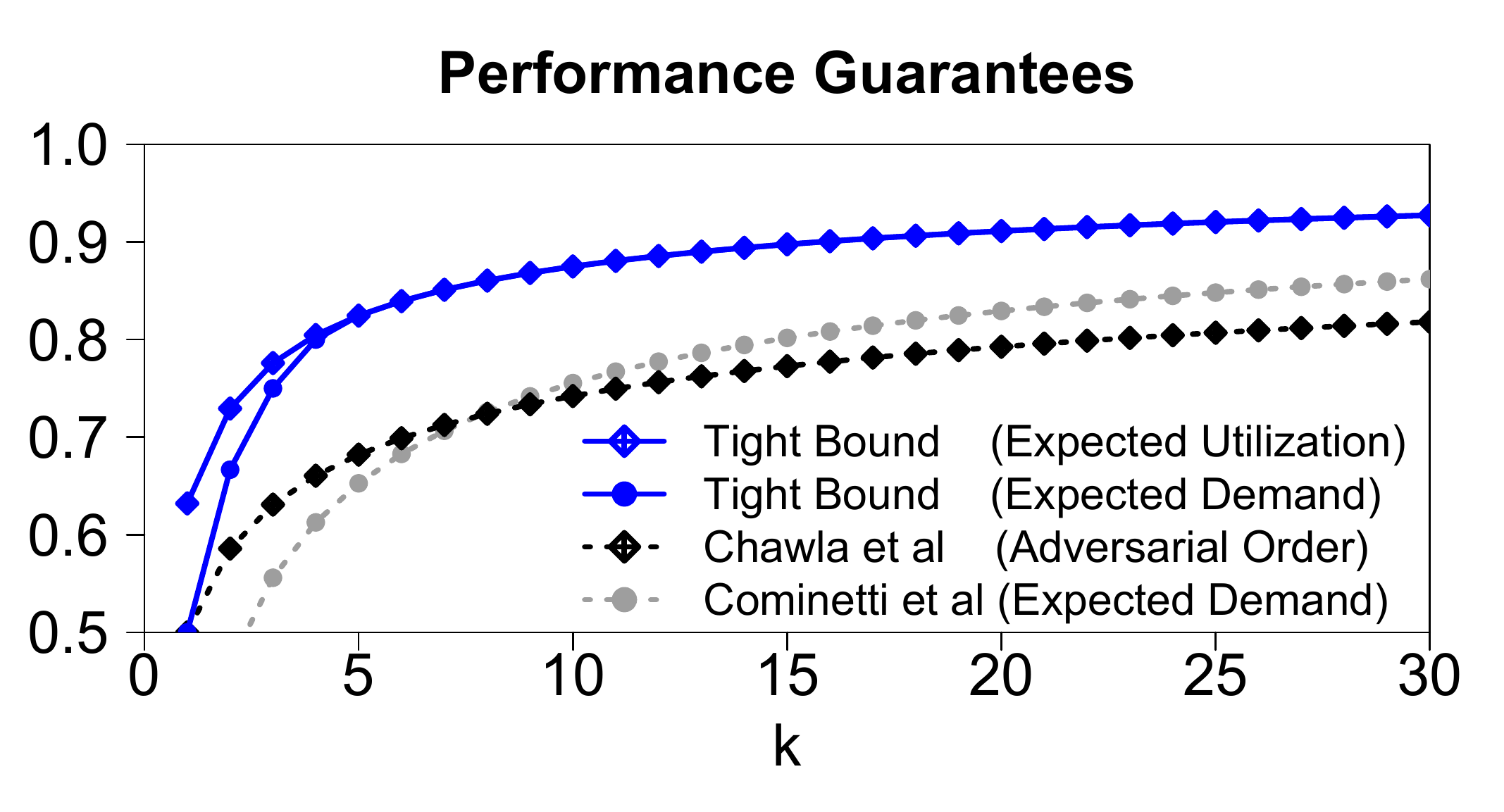}
\caption{Our tight bounds compared to bounds from prior work.
\citet{chawla-devanur-lykouris_2020} give lower bounds on the performance of static threshold policies when the arrival order is selected by an adversary. These bounds carry over to our setting, where the arrival order is random. In this setting, our Theorem \ref{thm:a-m} establishes that a policy based on expected utilization gives the best-possible guarantee.
In the random arrival model, \cite{cominetti2010optimal} provide lower bounds on the performance of a policy that sets expected demand equal to $k$. Our Theorem \ref{thm:a-lp} improves their result by giving a tight guarantee for this policy. 
}
\label{fig:comparison}
\end{figure}

It is insightful to compare $\gamma_k$ to guarantees from prior work. Figure \ref{fig:comparison} plots our guarantee against one proven by \cite{cominetti2010optimal}, as well as a guarantee of \cite{chawla-devanur-lykouris_2020} for the case where values arrive in an order selected by an adversary. In this latter model, it is known that no static threshold policy can achieve performance better than $1-\Theta(\sqrt{\log k/k})$ times the prophet's value. By contrast, Stirling's approximation implies that $\gamma_k\approx 1-1/\sqrt{2\pi k}$, so there is a separation of order $\sqrt{\log k}$ in the asymptotic loss between models with random and adversarial arrival order.

\subsection{Thresholds Based on Expected Demand} \label{sec:introExpD}

Jointly, Theorems \ref{thm:upper-bound} and \ref{thm:a-m} identify, for each $k$, the exact guarantee $\gamma_k$ that can be achieved by static threshold policies. However, it is natural to wonder about other ways to set the threshold. Do other simple policies also achieve this same guarantee?

One advantage of our analysis is that it only requires small modifications to answer this question. We focus on a simple policy which sets the threshold so that the expected number of values exceeding the threshold is equal to $k$. This policy is intuitive, and has previously been studied by \citet{cominetti2010optimal}, who show (under an assumption on the distributions $F_i$) that its performance is at least $1-\sqrt{\frac{1+1/\sqrt{k}}{2(k+1)}}$ times the prophet's value. \Cref{thm:a-lp} gives tight bounds for the performance of this policy which hold for any value distributions. Figure \ref{fig:comparison} offers a visual comparison between our tight bounds and those of \cite{cominetti2010optimal}.

It is intuitive that if expected demand is equal to $k$ and $k$ is large, then realized demand should be ``close" to $k$, implying near-optimal performance. By contrast, when $k$ is small, the following example shows that setting expected demand equal to $k$ does not guarantee strong performance.

\begin{example} \label{ex:a-lp-bad}
Fix $k$ and consider a small $\eps>0$.  Let there be $n=k+1$ applicants.  Applicants $i=1,\ldots,k$ all have a value that is distributed as: 1 with probability~$1-\eps/k$; 0 otherwise.
The value of applicant $k+1$ is distributed as: $1/\eps^2$ with probability~$\eps$; 0 otherwise.
\end{example}

On this example, the prophet's value is at least $1/\eps$, as this is the expected value of applicant $k+1$. Meanwhile, the policy that sets expected demand equal to $k$ will accept all applicants with non-zero values.  Therefore, if applicants $1,\ldots,k$ all have value $1$ and also arrive before applicant $k+1$, then the final applicant is rejected. By a union bound, the probability of this is at least $\frac{1-\eps}{k+1}$. Because the contribution from the first $k$ applicants is upper-bounded by $k$, the value of this policy is at most
$$\frac{1}{\eps^2}\cdot\eps(1-\frac{1-\eps}{k+1})+k = \frac{1}{\eps}\cdot\frac{k}{k+1}+\frac{1}{k+1} + k.$$ 
Taking $\eps\to0$, this shows that the policy which sets expected demand equal to $k$ does not guarantee more than $\frac{k}{k+1}$ times the prophet's value.

Examples \ref{eg:countereg} and \ref{ex:a-lp-bad} both provide upper-bounds on the performance of a threshold set so that expected demand equals $k$. These upper bounds are $\gamma_k$ and $k/(k+1)$ times the prophet's value, respectively. Our next result establishes that these examples represent worst cases for this policy.


\begin{theorem} \label{thm:a-lp}
For any instance $(k,\bF)$, the performance of any threshold such that expected demand is equal to $k$ is at least $\min\{\gamma_k,\frac{k}{k+1}\}$ times the prophet's value.
\end{theorem}

Note that $\min\{\gamma_k,\frac{k}{k+1}\}$ is $\frac{k}{k+1}$ when $k\le 4$, and $\gamma_k$ when $k>4$.
Therefore, Theorems \ref{thm:upper-bound} and \ref{thm:a-lp} jointly imply that the policy of setting expected demand equal to $k$ achieves the best-possible guarantee when $k>4$. 

In \Cref{ex:a-lp-bad}, we intuitively want expected demand to be below $k$, in order to increase the probability of accepting the high-value applicant. In fact, \cite{hajiaghay-kleinberg-sandholm_2007} analyze a policy that sets expected demand equal to $k - \sqrt{2k \log k}$. Unfortunately, we show in Proposition \ref{prop:expDemandBad} that when $k \le 4$, any algorithm that sets expected demand to some fixed target (even if that target is not $k$) \textit{cannot} guarantee $\gamma_k$ times the prophet's value. Thus, while expected utilization can be used to achieve optimal guarantees for all $k$, expected demand can only offer optimal guarantees for $k > 4$. However, Proposition \ref{prop:iid} establishes that if values are IID, then setting expected demand equal to $k$ offers a guarantee of $\gamma_k$ for all $k$.  We present these results in \Cref{sec:demandStatPolicies}.


\subsection{A New Result for Bernoulli Optimization}

Our analysis deploys a result about optimization problems involving sums of independent Bernoulli random variables. We highlight this result here, because we believe that it will be of general interest to researchers who  encounter Bernoulli optimization problems in other contexts.

Given a vector of probabilities ${\bf p} \in [0,1]^n$, let $\Dp$ denote the sum of independent Bernoulli random variables with means $p_1,\ldots,p_n$. For any functions $f,g$ on the non-negative integers, consider the problem of choosing ${\bf p}$ to minimize $\bE[f(\Dp)]$ subject to the constraint that $\bE[g(\Dp)] = \phi$:
\begin{align}
\Phi_n(f,g,\phi)= \min_{\vp \in[0,1]^n}& \,\, \bE[f(\Dp)] \label{eqn:Phi}
\\ \text{s.t.}\,\,\, & \,\,\bE[g(\Dp)] =\phi. \nonumber
\end{align}
We show that for {\em any} functions $f$ and $g$, this problem has a solution with a very simple structure.

\begin{theorem} \label{thm:structuralGeneric}
For any $n \in \mathbb{N},$ any $f, g: \mathbb{N}_0 \rightarrow \mathbb{R}$ and any $\phi \in \mathbb{R}$ such that \eqref{eqn:Phi} is feasible, there is a value $p \in (0,1)$ and an optimal solution $\vp$ to \eqref{eqn:Phi} such that every $p_i \in \{0, p, 1\}$.
\end{theorem}

This surprising\footnote{It seems natural that a symmetric function on a hypercube should either have a symmetric optimum or an optimum at the boundary of the feasible region. Theorem \ref{thm:structuralGeneric} confirms this intuition when the problem has a particular structure. However, this is not true in general: let $h(z) = \min((z+1)^2,(z-1)^2)$, and let $f(x,y) = x^2 + y^2 + h(x-y)+h(y-x)$. Then $f$ is symmetric, and its minima are $(x,y) \in \{(-0.5,0.5), (0.5,-0.5)\}$. } result says that it suffices to consider cases where $\Dp$ is equal to a constant plus a binomial random variable. Corollary 2.1 in \citet{hoeffding_1956} establishes this result for arbitrary $f$ when $g$ is the identity, and Lemma 7 of \citet{chawla-devanur-lykouris_2020} establishes this result when $f(d) = \min(1,d/k)$ and $g(d) = \bI(d < k)$. We believe that Theorem \ref{thm:structuralGeneric}, which generalizes these preceding results, will be useful in settings beyond the prophet secretary problem.


\subsection{Roadmap}

\Cref{sec:lit-review} provides a thorough review of secretary problems and prophet inequalities, and compares our results to prior work.
\Cref{sec:a-m-proof} provides formal definitions and proves \Cref{thm:a-m}.
\Cref{sec:a-lp-proof} provides the modifications necessary to prove \Cref{thm:a-lp}.
\Cref{sec:conc} discusses several extensions, including robustness of our policies when values are imperfectly observed, additional results for policies based on expected demand, and other rules for setting thresholds.
We defer the proofs of \Cref{thm:upper-bound} and \Cref{thm:structuralGeneric}, as well as all intermediate results, to the Appendix. 


\section{Literature Review} \label{sec:lit-review}

There is a large literature on sequential selection problems, dating back at least to \citet{cayley1875mathematical}. The name ``prophet secretary problem" was coined by \citet{esfandiari2017prophet}, and inspired by prior work on ``secretary problems" and ``prophet inequalities". Section \ref{sec:history} discusses the different modeling assumptions made by these two lines of work. Section \ref{sec:resultComparison} provides more detailed comparisons of existing results to our own. 


\subsection{Secretary Problems and Prophet Inequalities} \label{sec:history}



\textbf{The Secretary Problem.} The most famous secretary problem features a sequence of $n$ unknown values arriving in uniformly random order, with the objective being to select the maximum. According to \citet{ferguson1989solved}, this problem first appeared in print in a 1960 Scientific American column by Martin Gardner, but was widely known even before then. The optimal solution involves skipping the first $n/e$ values, and selecting the first value to exceed the highest of these \citep{dynkin1963optimum}. 

\textbf{Prophet inequalities.}
Attributed to  \citet{krengel1977semiamarts,krengel1978semiamarts}, ``prophet inequalities" commonly refer to comparisons between online and offline selection algorithms on a sequence of cardinal values drawn from known distributions.
\cite{samuel1984comparison} initiated the study of static thresholds by proving the elegant result that setting a threshold equal to the median of the maximum value collects at least 1/2 of this maximum value in expectation. 
More recently, \citet{kleinberg2012matroid} show that this same guarantee is achieved by setting a threshold equal to half of the expected maximum value.

\textbf{Extensions of these problems.} Many variants of these classic problems have been studied. In a single paper, \citep{gilbert1966recognizing} consider four: one in which multiple items can be selected, another in which distributional information about the values is available up front, a third in which values arrive in an adversarial order, and a fourth in which the payoff is not $0$ or $1$ but rather depends on the values themselves. Meanwhile, follow-up work on prophet inequalities has studied the ``free-order" setting where the decision-maker can choose the order of distributions \citep{hill1983prophet,hill1985selection,beyhaghi2020improved,agrawal2020optimal} and also settings where the distribution of values is unknown but samples from these distributions are available \citep{azar2014prophet,correa2019prophet,rubinstein2020optimal}.

\citet{ferguson1989solved} writes ``Since there are so many variations of the basic secretary problem...it is worthwhile to try to define what a secretary problem is." He concludes ``a secretary problem is a sequential observation and selection problem in which the payoff depends on the observations only through their relative ranks and not otherwise on their actual values."

However, many subsequent papers have followed a different convention, and used the term ``secretary'' to refer to problems where 
unknown values arrive in a uniformly random order. \citet{rubinstein2016beyond} draws the following distinction:
\begin{quote}
    \it In
the Secretary Problem, the values of items are chosen adversarially, but their arrival order is
random. In Prophet Inequality, the order is adversarial, but the values are drawn from known,
independent but not identical distributions.
\end{quote}

This terminology has been applied fairly consistently within the computer science literature. For example, \citet{ezra2020secretary} (``Secretary Matching with General Arrivals'') study matching models with adversarial values arriving in a random order, while \citet{ezra2021prophet} (``Prophet Matching with General Arrivals") consider the case where values are drawn from known distributions and arrive in an order fixed by an adversary. 

The ``multiple-choice secretary" model of \citet{kleinberg_2005} features $n$ adversarially selected values arriving in random order, as does later work on 
 the knapsack \citep{babaioff2007knapsack} and matroid \citep{babaioff2007matroids} secretary problems, 
secretary matching problem \citep{ezra2020secretary}, and secretary problem under general downward-closed feasibility constraints \citep{rubinstein2016beyond}.

Similarly, prophet inequalities have been studied under $k$-unit \citep{hajiaghay-kleinberg-sandholm_2007,alaei2014bayesian}, matroid \citep{kleinberg2012matroid}, knapsack \citep{dutting2020prophet}, matching \citep{ezra2021prophet}, and general downward-closed \citep{rubinstein2016beyond} feasibility constraints. All of these papers feature values drawn from known distributions, arriving in an order that may be selected by an adversary. This literature was surveyed by \cite{hill1992survey}, and more recent work is discussed in \citet{lucier2017economic,correa-foncea-hoeksma-oosterwijk-vredeveld_2018}.


Inspired by this convention, \cite{esfandiari2017prophet} use the phrase ``prophet secretary problem" to describe settings in which values are drawn from (possibly heterogeneous) known distributions and presented in a uniformly random order. We adopt this terminology, while noting that it is not used universally. For example,  \citet{arlotto2019uniformly} and \citet{bray2019does} use the term ``multisecretary" to refer to a problem where values are drawn IID from a known distribution. These papers study (additive) regret, in contrast to the multiplicative guarantees in our work.

\subsection{Comparison to Existing Results} \label{sec:resultComparison}

In this \namecref{sec:resultComparison} we compare our guarantees to those in prior work. \citet{kleinberg_2005} considers a ``multiple-choice secretary" model in which $n$ values arrive in random order, and the decision-maker can accept $k$ values and wishes to maximize their sum. He provides an algorithm that guarantees $1-5/\sqrt{k}$ times the prophet's value. Unlike his work, ours assumes that values are drawn independently from known distributions. This allows us to use simple static threshold policies, and to get a better guarantee of $\gamma_k \approx 1 - 1/\sqrt{2\pi k}$.

The remainder of this section compares to papers which assume that values are drawn from known distributions. Some of these papers assume values are IID; others assume they are drawn from heterogeneous distributions and arrive in a random order; others assume that the arrival order is chosen by an adversary\footnote{For exact definitions of the different levels of adversarial power in selecting the order, see \citet{feldman2021online}.}. Worst-case guarantees with an adversarial arrival order translate immediately to random arrival order, while guarantees for random arrival order translate to the IID case. This allows for a streamlined comparison of results, which is provided in \Cref{tab:my_label}. We now discuss the results in \Cref{tab:my_label}, along with some earlier related work.

\SingleSpacedXI
\begin{table}[]
\centering
\begin{tabular}{c|c|ccc} \hline
 & & IID & Random Arrival Order & Adversarial Arrival Order \\ 
\hline
 & $k=1$ & 0.745& (0.669,\ 0.732)  & 1/2 \\
& &  LB: \cite{correa_foncea_hoeksma_oosterwijk_vredeveld_2017}, & \cite{correa-saona-ziliotto_2020}  & \cite{krengel1978semiamarts} \\ 
General   & & UB: \cite{hill-kertz_1982} \\
\cline{2-5}
Online & $k>1$ & $\left(\alpha_k, 1 \right)$  & $\left(1-e^{-k}\frac{k^k}{k!}, 1 \right)$ & $\left(1 - \frac{1}{\sqrt{k+3}},1-\Omega\left(\frac{1}{\sqrt{k}}\right)\right)$ \\
Policies & & LB: \citet{beyhaghi2020improved} & \textbf{NEW}  & LB: \cite{alaei2014bayesian}, \\
& & & & UB: \cite{hajiaghay-kleinberg-sandholm_2007} \\
\hline
 & $k=1$ & $1-1/e$  & $1-1/e$ & 1/2 \\
Static & & \cite{ehsani2018prophet} &  \cite{ehsani2018prophet} &  \cite{samuel1984comparison} \\
\cline{2-5}
Threshold & $k>1$ &   $1-e^{-k}\frac{k^k}{k!}$ &  $1-e^{-k}\frac{k^k}{k!}$ & $\left(\beta_k, 1-\Omega{\sqrt{\frac{\log k}{k}}}\right)$ \\
Policies & & LB: \citet{yan2011mechanism}  & \textbf{NEW}  & LB: \cite{chawla-devanur-lykouris_2020}, \\
  & & UB: \textbf{NEW}{\color{white} filler}& & UB: \cite{ghosh-kleinberg_2016}  \\ \hline
\end{tabular}
\caption{A comparison of the fraction of the prophet's value that can be guaranteed in three different models: IID, random arrival order (``prophet secretary"), and adversarial arrival order. All guarantees are tight unless (lower, upper) bounds are indicated in parentheses. In this case, LB and UB refer to work establishing lower and upper bounds, respectively.  Our work establishes tight performance guarantees for static threshold policies in the random order model.
Our results also provide improved lower bounds for arbitrary policies. \\
The constants $\alpha_k$ are from \citet[Tbl.~3]{beyhaghi2020improved}, whose results for the free-order model also apply to IID distributions. 
The constants $\beta_k$ are from \citet{chawla-devanur-lykouris_2020}.
}
\label{tab:my_label}
\end{table}
\DoubleSpacedXI

\textbf{Prophet secretary with a single item.}
The prophet secretary problem with a single item was introduced in \citet{esfandiari2017prophet}, where it was shown that a general policy can guarantee $1-1/e\approx0.632$ times the prophet's value.
This guarantee was improved to $1-1/e+1/400$ in \citet{azar-chiplunkar-kaplan_2018}, and more recently to $0.669$ in \citet{correa-saona-ziliotto_2020}.
In the special case of IID values, a tight guarantee of 0.745 is known \citep{hill-kertz_1982,correa_foncea_hoeksma_oosterwijk_vredeveld_2017}.
To our understanding, none of these analyses for general policies are easy to extend to take advantage of having multiple items, and our guarantee of $1-e^{-k}\frac{k^k}{k!}$ is the best-known guarantee for $k>1$ among {\em all} policies.

\textbf{Static thresholds for prophet secretary.}
It was originally shown in \cite{esfandiari2017prophet} that a static threshold policy cannot guarantee more than 1/2 times the prophet's value.
However, \citet{ehsani2018prophet} later showed that the tight ratio actually increases to $1-1/e$ if one assumes continuous distributions, or allows for randomized tie-breaking.
We also allow randomized tie-breaking, and establish the tight guarantee of $1-e^{-k}\frac{k^k}{k!}$ for every $k$, generalizing the result of \citet{ehsani2018prophet} for $k=1$. We note, however, that our proof technique is very different from \citet{ehsani2018prophet}, who draw uniform arrival times from [0,1] for the agents.  The policy of \citet{ehsani2018prophet} for $k=1$ was also later analyzed without arrival times in \citet[Sec.~2]{correa-saona-ziliotto_2020}, who characterize the worst case using Schur-convexity.  It is unclear whether Schur-convexity can be used to prove our general Bernoulli optimization result (\Cref{thm:structuralGeneric}), because the constraint on the probability vector $\vp$ involves an \textit{arbitrary} function $g$.  Similarly, it is difficult to apply Schur-convexity given our $g$ function in \Cref{thm:a-m}, unless $k=1$.
Finally, Schur-convexity may not hold for our result in \Cref{thm:a-lp}, if $k<5$.

\citet{cominetti2010optimal} consider a related model where there are $k$ last-minute slots to offer to a set of customers with known values and acceptance probabilities. If more than $k$ customers accept the offer, then a randomly selected $k$ of these customers receive the slots. This is equivalent to a special case of the prophet secretary problem in which each $F_i$ places mass on two values (one of which is zero). They analyze a policy that makes offers such that in expectation, $k$ customers accept. Their Proposition 3 establishes a guarantee of $1-\sqrt{\frac{1+1/\sqrt{k}}{2(k+1)}}$ for this policy. Our \Cref{thm:a-lp} studies the same policy, improving 
their lower bound to  $\min\{\gamma_k,\frac{k}{k+1}\}$, and showing that this bound is best-possible when value distributions can be arbitrary.

\textbf{Prophet inequalities for IID distributions with $k$ items.}
Motivated by posted-price mechanisms,
\citet{yan2011mechanism} established the guarantee of $\gamma_k$ for IID values, and showed this to be the best-possible guarantee relative to the LP benchmark.
We extend this result by showing that the same guarantee holds when values are \textit{non-identically} distributed but arrive in a random order.
Furthermore, our \Cref{thm:upper-bound} shows that for static threshold policies, the same upper bound holds even when comparing against the weaker benchmark of prophet's value.

\textbf{Relationship with posted-price mechanisms.}
Static threshold policies in our sequential selection problem translate to deterministic posted-price mechanisms in a Bayesian auctions setting where buyers have valuations drawn IID from a \textit{regular} distribution (see \citet{correa2019pricing} and \citet{correa2021optimal}).
There, \citet{dutting2016revenue} have shown the existence of regular valuation distributions for which a deterministic posted price cannot earn more than $\frac{1}{k}\bE[\min\{\Bin(n,k/n),k\}]$ times the revenue of Myerson's optimal auction.
As $n\to\infty$, this bound approaches $\gamma_k$. Through the translation of \citet{correa2019pricing}, this implies the same upper bound as our \Cref{thm:upper-bound}.  We include \Cref{thm:upper-bound} because it is directly stated in the sequential selection setting, which allows for explicit constructions of the worst-case distributions and direct analysis of the optimal threshold.

\textbf{Adversarial-order prophet inequalities.}
Under adversarial arrival order, \citet{hajiaghay-kleinberg-sandholm_2007} originally showed that setting a static threshold so that expected demand equals $k-\sqrt{2k\log k}$ yields a guarantee of $1-O(\sqrt{\frac{\log k}{k}})$ for large $k$.  The asymptotic error term of order $\sqrt{\frac{\log k}{k}}$ was shown to be tight in \citet{ghosh-kleinberg_2016}.
Since $\gamma_k=1-\Theta(\sqrt{\frac{1}{k}})$, there is a separation between the asymptotic performance of static threshold policies in random and adversarial arrival models.

We should note that under adversarial order, a guarantee of order $1-\Theta(\sqrt{\frac{1}{k}})$ can be recovered if one goes beyond static threshold policies.  In fact, a sequence of papers \citep{alaei2011bayesian,alaei2012adallocation,alaei2014bayesian} establish a well-known guarantee of $1-\frac{1}{\sqrt{k+3}}$ for general policies that holds for all $k$.
This is asymptotically tight, due to an upper bound of $1-\Omega(\sqrt{\frac{1}{k}})$ from \citet{hajiaghay-kleinberg-sandholm_2007}. 
The guarantee of $1-\frac{1}{\sqrt{k+3}}$ was improved for small values of $k$ in \citet{chawla-devanur-lykouris_2020}, and recently a tight result for all $k$ was found in \citet{jiang2022tight} using a different technique.
Since we analyze static thresholds, our techniques build upon \citet{chawla-devanur-lykouris_2020}.

\textbf{Implications for online multi-resource allocation and contention resolution schemes.}
Our guarantee of $\gamma_k=1-e^{-k}\frac{k^k}{k!}$ holds (and is tight) relative to the LP relaxation, when allocating $k$ identical itemss and demand arrives in random order.
This guarantee directly translates to general online multi-resource allocation problems in which  there are at least $k$ copies of each resource and demand arrives in random order, by standard decomposition results \citep[see e.g.][]{alaei2014bayesian,gallego2015online}.
Relatedly, our tight ex-ante prophet inequality of $\gamma_k$ for random-order equivalently implies a tight $\gamma_k$-selectable random-order contention resolution scheme for all $k$-uniform matroids \citep{lee2018optimal}.  We defer further discussion of these implications to the cited papers.

\section{Proof of Theorem \ref{thm:a-m}} \label{sec:a-m-proof}

We begin by defining our notation and terminology. We denote the set of positive integers by $\mathbb{N}$, the set of non-negative integers by $\mathbb{N}_0$, and the set of real numbers by $\mathbb{R}$ and the set of non-negative real numbers by $\mathbb{R}_+$.
Given values $\bV \in \mathbb{R}_+^n$ and a threshold $t \in \mathbb{R}_+$, define
\begin{align}
D^t(\bV)=\sum_{i=1}^n\bI(V_i > t). \label{eq:demand}
\end{align}
We refer to this as ``demand" at threshold $t$. For $d \in \mathbb{N}_0$ and $k \in \mathbb{N}$ define
\begin{equation}\MR_k(d) = \min\left(1,\frac{d}{k}\right).\label{eq:match-rate}\end{equation}
The letters $\MR$ stand for ``utilization'': when realized demand is $d$, $\MR_k(d)$ gives the fraction of items that are allocated.

\begin{definition}
\it 
Given an instance defined by $k \in \mathbb{N}$ and distributions $\bF = \{F_i\}_{i = 1}^n$, 
\begin{itemize}
    \item The {\bf expected demand} of threshold $t\in \mathbb{R}_+$ is $\bE_{\bV \sim \bF}[D^t(\bV)] =  \sum_{i = 1}^{n} (1 - F_i(t))$.
    \item The {\bf expected utilization} of threshold $t \in \mathbb{R}_+$ is $\bE_{\bV \sim \bF}[\MR_k(D^t(\bV))]$.
    \item The {\bf performance} or {\bf value} of  threshold $t \in \mathbb{R}_+$ is 
    \begin{equation}\ST_k^t(\bF) = \bE_{\bV \sim \bF} \left[\sum_{i = 1}^n V_i \bI(V_i > t) \min\left(1,\frac{k}{D^t(\bV)}\right) \right].\label{eq:static-threshold}\end{equation}
    \item The {\bf prophet's value} is the expected sum of the $k$ largest values, and is denoted $\OMN_k(\bF)$.
    \item The {\bf LP relaxation}  or {\bf ex-ante value} is defined as
\begin{align*}
\LP_k(\bF)& =\sup\left\{\sum_{i=1}^n\int_{1-x_i}^1 F^{-1}(z)dz:\ \sum_{i=1}^n x_i \le k;\ x_i\ge0\ \forall i\right\} 
\end{align*}
\end{itemize}
\label{def:value}
\end{definition}

The performance of $t$ is equal to the expected sum of values of accepted applicants when using threshold $t$. This expectation is taken over the realized values $V_i$ and the arrival order of applicants. To see this, note that if we condition only on the values $V_i$ and take expectations over the arrival order, the probability that applicant $i$ is accepted is exactly $\bI(V_i > t)\min\left(1,\frac{k}{D^t(\bV)}\right)$. This is because if $D^t(\bV) \le k$, every applicant with $V_i>t$ is accepted with certainty, while if $D^t(\bV) > k$, then $k$ of the $D^t(\bV)$ applicants with values exceeding $t$ will be accepted.

Meanwhile, in the LP relaxation, $x_i$ denotes the probability of accepting $i$ and the constraint that at most $k$ applicants are accepted only needs to hold in expectation.
This relaxation accepts each applicant $i$ whenever $i$ takes a value in its top $x_i$'th quantile.

\subsection{Trading off Two Risks}

When choosing a threshold $t \in \mathbb{R}_+$, there is a tradeoff. If $t$ is too high, utilization will be low. Conversely, if $t$ is too low, a high-value applicant may have a low probability of being accepted. We capture this second risk using the following function. 
For $k\in\mathbb{N}$ and $d \in \mathbb{N}_0$, define
\begin{align}
\AR_k(d) &=\min\left(1,\frac{k}{d+1} \right). \label{eq:acceptance-rate}
\end{align}
The letters $\AR$ stand for ``acceptance rate'': because applicants arrive in random order, an applicant who competes with $d$ others for $k$ items will be accepted with probability $\AR_k(d)$.  
Intuitively, the function $\MR$ describes the risk of under-allocation, while $\AR$ describes the risk of allocating items too quickly.  Our next result uses these functions to bound the performance of any static threshold. 

\begin{lemma} \label{lem:lb}
For all $k \in \mathbb{N}$, $t\in \mathbb{R}_+$ and all $\bF$, 
\[ \frac{\ST_k^t(\bF)}{\OMN_k(\bF)} \geq \frac{\ST_k^t(\bF)}{\LP_k(\bF)} \geq \min\left(\mathop{\bE}_{\bV \sim \bF} \left[\MR_k(D^t(\bV)) \right],\mathop{\bE}_{\bV \sim \bF} \left[\AR_k(D^t(\bV)) \right]\right).\]
\end{lemma}
This result is analogous to Lemma 1 in \citet{chawla-devanur-lykouris_2020}. However, they assume a fixed arrival order selected by an adversary. In that model, the risk from allocating items too quickly is higher than in our random-arrival model. Correspondingly, their result replaces $\mathop{\bE}_{\bV \sim \bF} \left[\AR_k(D^t(\bV)) \right]$ with the lower quantity $\bE_{\bV \sim \bF}[\bI(D^t(\bV) < k)]$. This distinction is the source of our improved bounds shown in Figure \ref{fig:comparison}.

\subsection{Bernoulli Optimization}

Note that both terms in the lower bound in Lemma \ref{lem:lb} depend on $\bF$ only through the parameters $p_i= 1 - F_i(t)$. We now study the problem of choosing probabilities $p_i$ to minimize this lower bound.

For any positive integer $n$ and  $\vp=(p_1,\ldots,p_n)\in[0,1]^n$, let $\Dp$ denote the sum of independent Bernoulli random variables with means $p_1,\ldots,p_n$ (this is often referred to as the ``Poisson Binomial'' distribution). The problem of minimizing $\bE[\AR_k(\Dp)]$ subject to $\bE[\MR_k(\Dp)] = \gamma_k$ is a special case of the optimization problem in \eqref{eqn:Phi}, and thus Theorem \ref{thm:structuralGeneric} applies. In fact, 
our next result establishes the stronger conclusion that this problem has an optimal solution in which demand follows a binomial distribution.
\begin{lemma} \label{lem:structuralForMedian}
For all $k, n \in\mathbb{N}$, with $n > k$, and all $\phi \in (0,1)$, the optimization problem $\Phi_n(\AR_k,\MR_k,\phi)$ has an optimal solution in which all $p_i$ are equal.
\end{lemma}

This result is analogous to Lemma 11 in \cite{chawla-devanur-lykouris_2020}, except that they work with the function $f(d) = \bI(d < k)$ instead of $\AR_k$. 
We follow the proof technique from \citet{chawla-devanur-lykouris_2020} of optimizing two variables while holding the others fixed. This requires some case analysis, provided in \Cref{lem:two-opt} in Appendix \ref{sec:two-opt}.
Our treatment is more abstract than in \citet{chawla-devanur-lykouris_2020}, which leads to our more general result about Bernoulli optimization in \Cref{thm:structuralGeneric}.
Our proof of \Cref{lem:structuralForMedian} also requires non-trivial applications of classical facts about Poisson Binomial distributions, provided in  \Cref{lem:poisson-binomial} in Appendix \ref{sec:two-opt}.

\subsection{Completing the Proof} \label{sec:expUtilCompleting}

\text{ }

\vspace{-.3 in} Lemma \ref{lem:lb} implies that if $t$ is a threshold such that expected utilization equals $\gamma_k$, then
\begin{equation} \frac{\ST_k^t(\bF)}{\OMN_k(\bF)}\ge\frac{\ST_k^t(\bF)}{\LP_k(\bF)} \geq \Phi_n(\AR_k,\MR_k,\gamma_k).\label{eq:first-step-thm2} \end{equation}

Lemma \ref{lem:structuralForMedian} implies that
\begin{equation}\Phi_n(\AR_k,\MR_k,\gamma_k) = \bE[\AR_k(\Bin(n,p_{n,k}))],\label{eq:step2} \end{equation}
were $p_{n,k}$ is defined as the solution to 
\begin{equation} \bE[\MR_k(\Bin(n,p_{n,k}))] = \gamma_k.\label{eq:pn-def} \end{equation}

Note that $\Phi_n$ is weakly decreasing in $n$, because any feasible solution with $n$ values is also feasible with $n' > n$ (simply set $p_i = 0$ for $i > n$). Therefore, it follows from \eqref{eq:first-step-thm2} and \eqref{eq:step2} that
\begin{align}
\frac{\ST_k^t(\bF)}{\OMN_k(\bF)}\ge\frac{\ST_k^t(\bF)}{\LP_k(\bF)}  & \geq \lim_{n \rightarrow \infty} \bE[\AR_k(\Bin(n,p_{n,k}))]. \label{eq:combine-1}
\end{align}

Furthermore, the following result establishes that $p_{n,k} \le k/n$.
\begin{lemma}
For $1 \le k \le n$, define  $p_{n,k}$ to be the solution to \eqref{eq:pn-def}.
\label{lem:pn-ub}
Then $p_{n,k} \le k/n$.
\end{lemma}
Because $\AR_k$ is weakly decreasing, this implies that
\begin{equation}\bE[\AR_k(\Bin(n,p_{n,k}))] \geq \bE[\AR_k(\Bin(n,k/n))]. \label{eq:pn-ub}\end{equation} 

Le Cam's theorem on the convergence of the binomial to the Poisson implies that 
\begin{align}
\lim_{n \rightarrow \infty} \bE[\AR_k(\Bin(n,k/n))] &  = \bE[\AR_k(\Pois(k))], \label{eq:lecam}
\end{align}
and Lemma \ref{lem:poisson-facts} in Appendix \ref{app:thm1} states that 
\begin{equation} \bE[\AR_k(\Pois(k))] = \gamma_k. \label{eq:equals-gamma} \end{equation}
Combining \eqref{eq:combine-1}, \eqref{eq:pn-ub}, \eqref{eq:lecam}, and \eqref{eq:equals-gamma} completes the proof of Theorem \ref{thm:a-m}.



\section{Proof of Theorem \ref{thm:a-lp}} \label{sec:a-lp-proof}

Our earlier analysis used Lemma \ref{lem:lb} to reduce to the problem of minimizing $\mathop{\bE}_{\bV \sim \bF} \left[\AR_k(D^{t}(\bV)) \right]$ subject to the constraint that expected utilization equals $\gamma_k$. We now perform a similar analysis for thresholds such that expected demand equals $k$.

\begin{lemma} \label{lem:LPPerfBound}
For any instance $(k, \bF)$, if $t$ is such that 
$\mathop{\bE}_{\bV \sim \bF} \left[D^{t}(\bV)\right] = k$, then
\[\mathop{\bE}_{\bV \sim \bF} \left[\MR_k(D^{t}(\bV)) \right] \geq \mathop{\bE}_{\bV \sim \bF} \left[\AR_k(D^{t}(\bV)) \right].\]
\end{lemma}

Combining this result with Lemma \ref{lem:lb}, it follows that if $t$ is such that $\mathop{\bE}_{\bV \sim \bF} \left[D^{t}(\bV)\right] = k$, then
\begin{equation} \frac{\ST_k^t(\bF)}{\OMN_k(\bF)} \geq \Phi_n(\AR_k,\ID,k) 
\ge \lim_{n\to\infty}\Phi_n(\AR_k,\ID,k),\end{equation}
where $\ID$ is the identity function $\ID(d) = d$ and the second inequality follows because $\Phi_n$ is weakly decreasing in $n$, as noted in Section \ref{sec:a-m-proof}.

\begin{lemma} \label{lem:structuralForLP}
For all $k,n \in \mathbb{N}$ with $n > k$, the optimization problem for $\Phi_n(\AR_k,\ID,k)$ has an optimal solution in which every nonzero $p_i$ is equal.
\end{lemma}
Note that this conclusion is similar to, but not the same as, that in Lemma \ref{lem:structuralForMedian}. When $g = \MR_k$, {\em all} $p_i$'s could be assumed to be equal, whereas when $g = \ID$, an optimal solution may have some $p_i$'s equal to zero. The proof of this result is in Appendix \ref{sec:structuralForLP}. It invokes \Cref{thm:structuralGeneric}, as well as additional facts about Poisson Binomial random variables presented in \Cref{lem:poisson-binomial}. Lemma \ref{lem:structuralForLP} implies that
\begin{equation} \lim_{n\to\infty}\Phi_n(\AR_k,\ID,k)
=\inf_{n\ge k}\bE\left[\AR_k(\Bin(n,k/n))\right]. 
\end{equation}

Note that when $n = k$, the expression on the right is $k/(k+1)$, and as $n \rightarrow \infty$, Le Cam's theorem implies that it approaches $\bE[\AR_k(\Pois(k))]$, which is equal to $\gamma_k$ by Lemma \ref{lem:poisson-facts}. Therefore, to complete the proof of Theorem \ref{thm:a-lp}, all that remains is to show that either $n = k$ or $n \rightarrow \infty$ is the worst case.

\begin{lemma}
For $k \in \mathbb{N}$,
\begin{align} \label{eqn:secLPGoal2}
\inf_{n\ge k}\bE[\AR_k(\Bin(n,k/n))]= \min\left( \frac{k}{k+1},\gamma_k\right).
\end{align}
\label{lem:annoying}
\end{lemma}

Our proof of this result involves an intricate case decomposition over all values of $k$ and $n$, aided by numerical verification, as we now explain.

\begin{figure}
\centering
\includegraphics[width=0.48\textwidth]{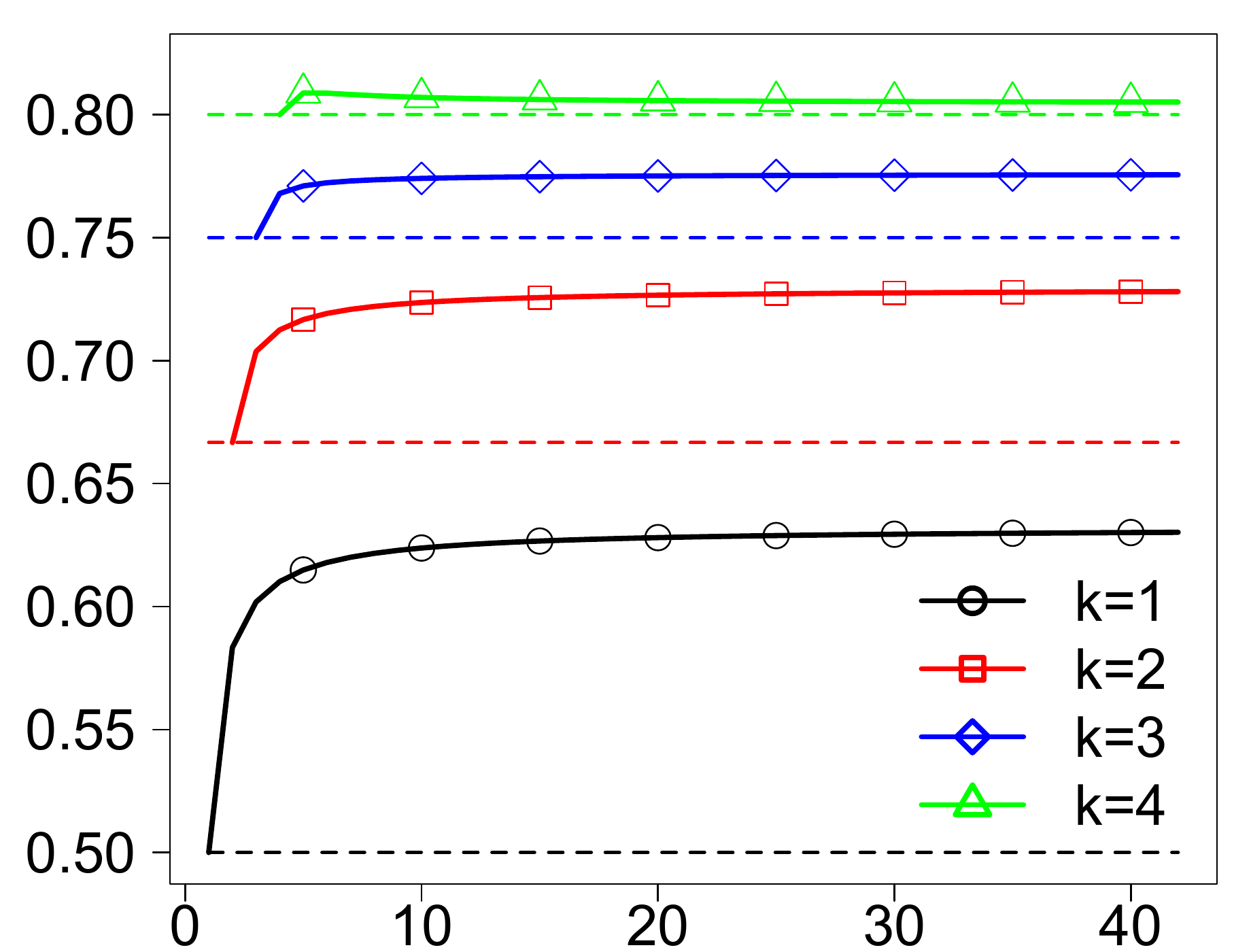}  \includegraphics[width=0.48\textwidth]{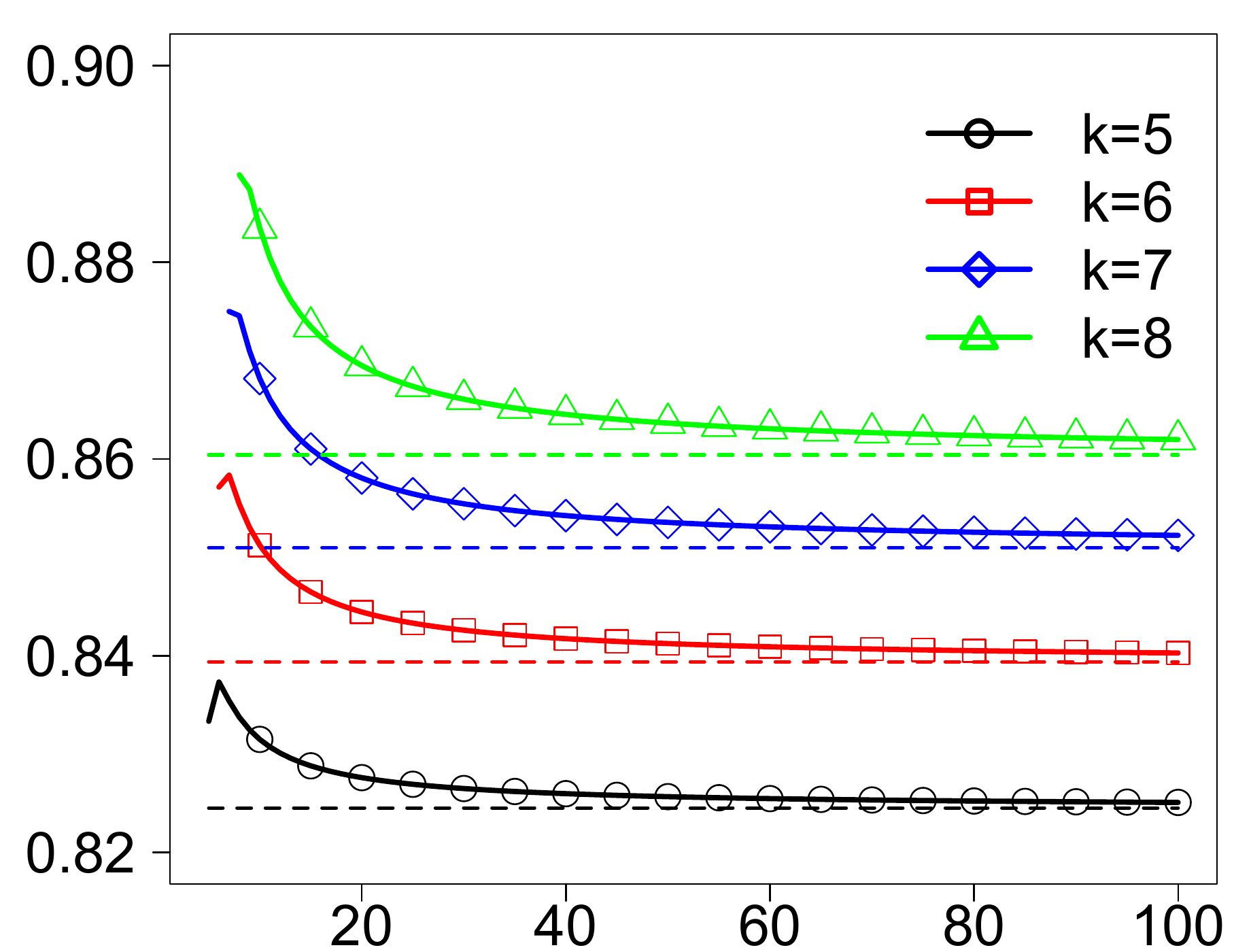}
\caption{Value of the expression on the left of \eqref{eqn:secLPGoal2}, for varying $k$ and $n$ (along $x$ axis). Dashed lines represent the lower bound $\min\left(\frac{k}{k+1},\gamma_k \right)$.
When $k \leq 4$ (left graph), the minimum value is $\frac{k}{k+1}$, and is obtained when $n=k$. When $k > 4$ (right graph), the minimum value is $\gamma_k$, and is obtained as $n \rightarrow \infty$.}
\label{fig:binomial-graphs}
\end{figure}

We first prove Lemma \ref{lem:annoying} for $k \le 4$. In this case, $k/(k+1) \le \gamma_k$. By Le Cam's theorem, $\bE[\AR_k(\Bin(n,k/n))] \geq \gamma_k - 2k/n^2$. This is larger than $k/(k+1)$ for $n \geq 42$. Figure \ref{fig:binomial-graphs} verifies that $\bE[\AR_k(\Bin(n,k/n))] \geq k/(k+1)$ for $n < 42$, completing the proof for $k \leq 4$. 

For $k \in \{5,6,7,8\}$, Figure \ref{fig:binomial-graphs} shows that $\bE[\AR_k(\Bin(n,k/n))]$ is minimized as $n \rightarrow \infty$, establishing the guarantee of $\gamma_k$.  Meanwhile, Lemmas \ref{lem:k31} and \ref{lem:k8} in Appendix \ref{sec:pfsLP} establish Lemma \ref{lem:annoying} for $k > 8$.

\section{Discussion: Robustness to Limited Information for a General Class of Demand Statistic Policies
} \label{sec:conc}

Our results precisely characterize the performance of static threshold policies in the prophet secretary problem, and provide the best-known guarantees for any online policy when $k > 1$. In addition, we show that simple policies -- such as trying to match the size of the eligible population to the number of available items -- achieve optimal guarantees. These guarantees are large enough to be reassuring in practice. For example, for a vaccination clinic with $k = 100$ doses, note that $\gamma_{100} > 96\%$. 

Although our work assumes that the policymaker can observe values directly and knows their distribution in advance, our algorithms and analysis in fact require much less information. We elaborate on this point below, and explain several associated benefits. First, our policies can be implemented even when the decision-maker does not directly observe applicants' values. Second, our analysis can offer guarantees even when only demand or utilization at {\em a particular} threshold are known (rather than the full demand distribution). Finally, our techniques could be readily adapted to analyze a broader class of policies for setting thresholds.


$$
$$

\vspace{-.2 in}
\subsection{Robustness to Monotone Transformations and Noisy Observations} \label{sec:monTransform}

In practice, it may be difficult to determine the exact value of allocating to each applicant. One advantage of our policies is that they only require the policymaker to determine which applicants have the highest value. They do not require knowledge of {\em how much} higher one applicant's value is than another. This is useful in settings where only a monotone transformation of values (rather than the values themselves) can be observed.

We illustrate this point using the application of COVID-19 vaccination. If the policymaker's goal is to minimize the expected number of deaths, then we can think of the value of vaccinating an individual as their pre-vaccination mortality risk.\footnote{This interpretation implicitly assumes that the vaccine is equally effective on all individuals, and that the policymaker's vaccination decisions do not affect the probability that unvaccinated individuals become infected. Nonetheless, allocating to individuals with the highest risk of death from infection is an intuitive and common approach.} However, mortality risk is not observed, and may be difficult to estimate. Instead, policymakers use proxies such as age and underlying medical conditions. 

Suppose that the policymaker plans to use an age-based eligibility rule, and that mortality risk is an increasing function of age. The {\em optimal} eligibility threshold depends on the exact relationship between age and mortality risk. If only the very elderly face a significant risk of death, then it makes sense to restrict eligibility to this population, even if that means that some doses go to waste. Conversely, if the risk of death is similar for all ages, then eligibility should be expanded, even if that means that some middle-aged people may receive the vaccine before the elderly population is fully vaccinated.

By contrast, so long as mortality risk increases with age, the eligibility thresholds selected by our policies do not depend on the exact relationship between the two. Instead, our policies can be implemented using only the distribution of applicants' ages, which may be much easier to estimate than the distribution of mortality risk.


For simplicity, the discussion above focused on the single variable of age. However, our insights also apply in a more general setting in which we do not directly observe $V_i$, but rather observe a type $\tau_i$, which may include age, co-morbidities, and other available information. 

The observables $\tau_i$ may insufficient to determine the true value of $V_i$. However, so long as there is a scoring function $s$ that maps types to real numbers and has the property that higher-scoring applicants have higher expected values (that is, $s(\tau_i) \ge s(\tau_j) \Leftrightarrow \bE[V_i \vert \tau_i] \ge \bE[V_i \vert \tau_i]$), then setting a threshold score for eligibility according to our policies will guarantee $\gamma_k$ times the value of choosing the $k$ applicants with the highest expected value conditioned on observables.




\subsection{Guarantees for Other Thresholds, and General Demand Statistic Policies} \label{sec:demandStatPolicies}

This paper focused on two particular policies, but a similar analysis could be deployed for analyzing other ways to determine the threshold.

\begin{figure}
\centerline{\includegraphics[width =.8  \textwidth]{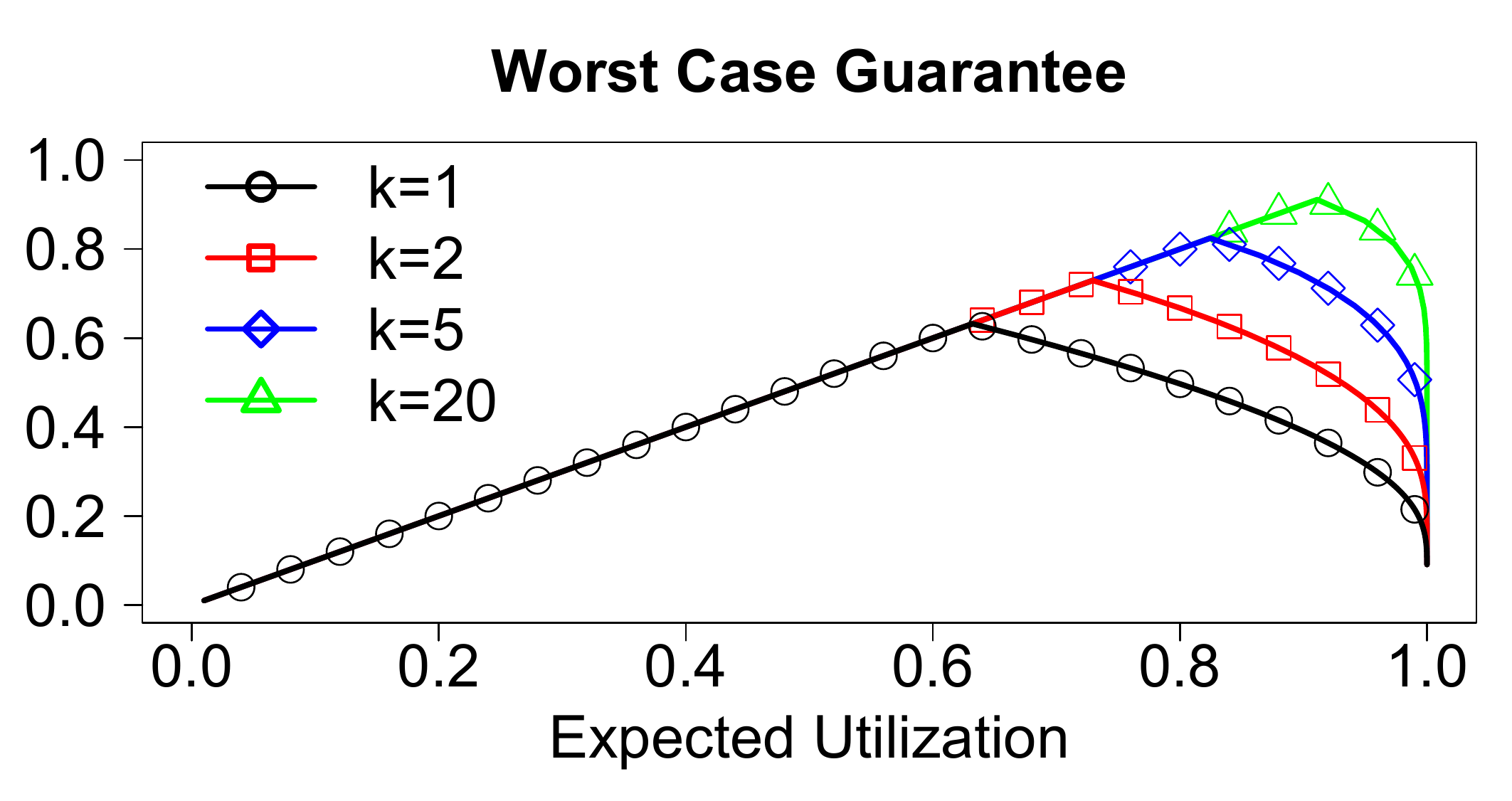}}
\caption{For any level of expected utilization achieved by a static threshold policy, our results imply a tight lower bound on the fraction of the prophet's value obtained, shown above. 
For each $k$, this lower bound ($y$-axis) is maximized and equals $\gamma_k$ when the expected utilization ($x$-axis) equals $\gamma_k$.  } \label{fig:ut}
\end{figure}

For example, our results immediately imply that for $k >  4$, any threshold in between the thresholds that we study also guarantees $\gamma_k$ times the prophet's value.
\footnote{The argument is simple. Define $a(t) = \bE[\MR_k(D^t)]$ and $b(t) = \bE[AR_k(D^t)]$. Because $a(t)$ is weakly increasing and $b(t)$ is weakly decreasing, it follows that if $\min(a(t_1),b(t_1)) \ge \gamma$ and $\min(a(t_2), b(t_2)) \ge \gamma$, then for any $t$ between $t_1$ and $t_2$ we have $a(t) \ge \min(a(t_1),a(t_2)) \ge \gamma$ and $b(t) \ge \min(b(t_1),b(t_2)) \ge \gamma$.
Therefore, by \Cref{lem:lb}, $\frac{\ST_k^t(\bF)}{\OMN_k(\bF)}\ge\gamma$.} In addition, \Cref{lem:lb} can be used to provide guarantees for thresholds outside of this range. For example, if a particular threshold has been used repeatedly, then expected utilization {\em at that threshold} is easy to estimate from past data.
Based on this expected utilization (and no other information about the value distributions $F_i$) we can use Lemmas \ref{lem:lb} and \ref{lem:structuralForMedian} to obtain the performance bounds in Figure \ref{fig:ut}. 

Perhaps even more interestingly, our approach could be used to analyze a general class of \textit{demand statistic policies}. These policies are specified by a non-decreasing function $g:\mathbb{N}\to\bR$ and a constant $\phi$, and choose the threshold $t$ so that $\bE[g(D^t)]=\phi$. 
In this paper, we focused on two such policies: one based on expected utilization with $g(d)=\min\{d/k,1\}$ and $\phi = \gamma_k$, and another based on expected demand with $g(d) = d$ and $\phi = k$.


All demand statistic policies inherit the advantage discussed in Section \ref{sec:monTransform}: they are robust to monotone transformations of values, and can be used when only proxies for value are available. Additionally, many organizations are already used to tracking statistics such as demand, utilization, and stockout frequency, and using these statistics to adjust their policies.

For these reasons, developing performance guarantees based on other demand statistics is a worthwhile direction for future work. Our Theorem \ref{thm:structuralGeneric} for Bernoulli optimization provides a useful starting point, since it simplifies the search for a worst-case demand distribution. One natural question is, which demand statistics $g$ are sufficient to deliver optimal guarantees, when paired with an appropriate choice of $\phi$?
Earlier we showed that expected utilization is sufficient, and expected demand is sufficient when $k>4$.
We now show that expected demand is \textit{insufficient} as a demand statistic when $k\le 4$, for any choice of $\phi$.


\begin{proposition}\label{prop:expDemandBad}
For $k<5$ and any value of $\phi$, there exists an instance $(k,\bF)$ such that the performance of any $t$ such that $\bE_{\bV \sim \bF}[D^t(\bV)] = \phi$ is strictly less than $\gamma_k$ times the prophet's value.
\end{proposition}

\Cref{prop:expDemandBad} is established in two cases.  If $d<k$, then the policy fails on an instance where all valuations are deterministically equal to 1, in which case it accepts too few applicants. On the other hand, if $d>k$, then the adversary can create an instance where all values are zero, except for a single applicant who has an infinitesimal probability of having a non-zero value.  In this case, the algorithm with target demand of $d$ accepts too many zero applicants, and risks being unable to accept the positive value when it occurs.

Despite this negative result, we note that setting expected demand equal to $k$ {\em does} provide an optimal guarantee if values are identically distributed (even for $k \le 4$).

\begin{proposition} \label{prop:iid}
If values are IID, then for any $k \ge 1$, the performance of any threshold such that expected demand is equal to $k$ is at least $\gamma_k$ times the prophet's value.
\end{proposition}

\textbf{Open question about another demand statistic.}
Another natural choice of demand statistic is $g(d)=\bI(d\ge k)$, in which case $\bE[g(D^t)]$ gives the ``stockout probability" (probability of running out of items). When $k = 1$, \cite{ehsani2018prophet} show that setting the stockout probability to $1 - 1/e$ provides an optimal guarantee of $\gamma_1 = 1 - 1/e$. In fact, when $k = 1$, stockout probability is equivalent to expected utilization, so their algorithm coincides with our own. We propose the following generalization of their algorithm for $k > 1$: set the threshold such that the stockout probability is equal to $\mathbb{P}(\Pois(k) \ge k)$. For large $k$, this means that the stockout probability is just slightly above $1/2$. This is a simple policy to explain and implement, and we conjecture that it also guarantees a $\gamma_k$ fraction of the prophet's value. 

\bibliographystyle{informs2014} 
\bibliography{bibliography}

\newpage

\begin{APPENDICES}

\section{Proof of Theorem \ref{thm:upper-bound}} \label{app:thm1}

We begin with a preliminary result about Poisson random variables.

\begin{lemma} \label{lem:poisson-facts}
For any $k \in \mathbb{N}$ and $\lambda\geq 0$, we have 
\begin{align}k \bE[\MR_k(\Pois(\lambda))] = \lambda \Pr(\Pois(\lambda)< k) + k\Pr(\Pois(\lambda) > k) = \lambda \bE[\AR_k(\Pois(\lambda))]. \label{eq:first-claim}\end{align}
Therefore, \begin{equation}  \bE[\MR_k(\Pois(k))] =
\bE[\AR_k(\Pois(k))] = \gamma_k. \label{eq:gamma-identity}\end{equation} 
\end{lemma}

\begin{myproof}[Proof of \Cref{lem:poisson-facts}]
For any random variable $D$, we have
\begin{equation}k \bE[\MR_k(D)] = \bE[ \min(D,k)] = k \Pr(D > k) + \bE[D \bI(D \leq k)].\label{eq:breakdown} \end{equation}
When $D$ is Poisson with mean $\lambda$,
\begin{equation}\bE[D \bI(D \leq k)] = \sum_{j = 0}^{k} j \frac{e^{-\lambda}\lambda^j}{j!} = \lambda \Pr(\Pois(\lambda)<k). \label{eq:pois-eq}\end{equation}
Combining \eqref{eq:breakdown} and \eqref{eq:pois-eq} yields the first equality in \eqref{eq:first-claim}. Meanwhile, for any $\lambda \ge 0$,
\begin{align*}
\bE[\AR_k(\Pois(\lambda))]& =\bE\left[\min\left(\frac{k}{\Pois(\lambda)+1},1\right)\right] \\
&=\Pr(\Pois(\lambda)<k)+\sum_{j=k}^{\infty}\frac{k}{1+j}e^{-\lambda}\frac{\lambda^j}{j!}
\\ &=\Pr(\Pois(\lambda)<k)+\frac{k}{\lambda}\sum_{j=k}^{\infty}e^{-\lambda}\frac{\lambda^{j+1}}{(1+j)!}
\\ &=\Pr(\Pois(\lambda)<k)+\frac{k}{\lambda}\Pr(\Pois(\lambda)>k).
\end{align*}
This establishes the second equality in \eqref{eq:first-claim}. Substituting $\lambda = k$ into \eqref{eq:first-claim} immediately yields \eqref{eq:gamma-identity}:
\[ \bE[\MR_k(\Pois(k))] = \bE[\AR_k(\Pois(k))] =  \Pr(\Pois(k)>k) + \Pr(\Pois(k)<k) = 1 - \Pr(\Pois(k)=k) =  \gamma_k.\]
\end{myproof}

We now prove Theorem  \ref{thm:upper-bound}, which follows immediately from Lemmas \ref{lem:counteregLB} and \ref{lem:counteregUB}. For $k \in \mathbb{N}$ define
\begin{align}
W_k & = k\cdot \frac{\Pr(\Pois(k)<k)}{\Pr(\Pois(k)>k)}.\label{eq:wk} 
\end{align}

\begin{lemma} \label{lem:counteregLB}
On Example~\ref{eg:countereg}, the prophet's value is at least $k+W_k-\frac{1+W_k}{n+1}$.
\end{lemma}
\begin{myproof}[Proof of \Cref{lem:counteregLB}]
If there are no high-value applicants, the prophet earns a reward of $k$. If there is at least one high-value applicant, then the prophet earns a reward that is at least $k+nW_k-1$. This implies that the prophet's expected reward is at least
\begin{align*}
k+(nW_k-1)\left(1-\left(1-\frac{1}{n^2}\right)^n\right)
& \geq k+(nW_k-1)\left(1 - \frac{1}{1 + 1/n}\right) \\
& = k+W_k-\frac{1+W_k}{n+1}.
\end{align*}
The inequality uses the fact that for $p \in [0,1]$ and $n \in \mathbb{N}$, $(1+pn)(1 - p)^n \leq 1$.
\end{myproof}
\begin{lemma} \label{lem:counteregUB}
On Example~\ref{eg:countereg}, the value of any static threshold algorithm is at most $$\gamma_k(k+W_k)+2kW_k(n^{-2/3} + n^{-1/3}).$$
\end{lemma}

Our proof uses the fact that if $N$ is a non-negative integer-valued random variable, then 
 \begin{equation}
\bE[N] = \sum_{j = 0}^\infty \Pr(N > j). \label{eq:favorite-id}
\end{equation}

\begin{myproof}[Proof of \Cref{lem:counteregUB}]
Clearly any algorithm should accept the high value of $nW_k$,
so we can restrict our search to fixed-threshold algorithms parameterized by a probability $p\in[0,1]$ of accepting the lower value of 1. Then each of the $n$ agents is accepted by the threshold independently with probability $q =  1/n^2 + (1 - 1/n)^2p$, and the expected number of accepted agents is
$\bE[\min\{\Bin(n,q),k\}]$.

Conditional on an agent being accepted, the expectation of his or her value is
\begin{align*}
\frac{(1-1/n^2)p}{1/n^2+(1-1/n^2)p}\cdot1 + \frac{1/n^2}{1/n^2+(1-1/n^2)p}\cdot nW_k\le 1+\frac{W_k}{nq}.
\end{align*}
Therefore, the expected value collected by any fixed-threshold algorithm is at most
\begin{align} \label{eq:FTAopt}
\max_{q\in[1/n^2,1]}\bE[\min\{\Bin(n,q),k\}]\left(1+\frac{W_k}{nq}\right).
\end{align}

We now consider two cases: either $q>n^{-2/3}$ or $q\le n^{-2/3}$.
If $q>n^{-2/3}$, then $nq>n^{1/3}$ and hence the value of~\eqref{eq:FTAopt} is less than $k(W_kn^{-1/3}+1)=k+kW_k n^{-1/3}$. Furthermore, $\gamma_k > 1/2$ and $W_k \geq k$ implies that $\gamma_k(k+W_k) > k$. \na{Maybe show $W_k > k$?}

In the other case where $q\le n^{-2/3}$, we note that
\begin{align*}
\bE[\min\{\Bin(n,q),k\}]
&=\sum_{j=0}^{\infty}\Pr[\Bin(n,q)=j]\min\{j,k\}
\\ &\le\sum_{j=0}^{\infty}\left(\Pr[\Pois(nq)=j]+\left|\Pr[\Bin(n,q)=j]-\Pr[\Pois(nq)=j]\right|\right)\min\{j,k\}
\\ &\le\bE[\min(\Pois(nq),k)]+2knq^2,
\end{align*}
where we have used Le Cam's theorem to bound the total variation distance between a $\Bin(n,q)$ random variable and a $\Pois(nq)$ random variable.
We conclude that the value of~\eqref{eq:FTAopt} is at most 
\begin{align*}
\bE[\min(\Pois(nq),k)]\left(1+\frac{W_k}{nq}\right)+2kW_kn^{-2/3}+2kn^{-1/3}.
\end{align*}
We will show that the first term in this expression is maximized when $nq = k$. When $nq = k$, Lemma \ref{lem:poisson-facts} implies that this term is equal to 
\[\bE[\MR_k(\Pois(k))]\left(k+W_k\right) = \gamma_k(k+W_k).\]

All that remains is to prove that the function $\bE[\min(\Pois(\lambda),k)]\left(\frac{W_k}{\lambda}+1\right)$ is maximized at $\lambda=k$. We first note that
\begin{align}
    \frac{d}{d\lambda} \bE[\min(\Pois(\lambda),k)] & = \Pr(\Pois(\lambda) < k). \label{eq:deriv}
\end{align}
This follows from the following calculations (the first of which uses \eqref{eq:favorite-id}):  
\begin{align}
\bE[\min(\Pois(\lambda),k)] &= \sum_{j = 0}^{k-1} \Pr(\Pois(\lambda)>j) = k - \sum_{j = 0}^{k-1}\Pr(\Pois(\lambda)\leq j). \label{eq:rewriting-enr}\\
\frac{d}{d\lambda} \Pr(\Pois(\lambda)\leq j) & = \frac{d}{d\lambda} \sum_{i = 0}^j \frac{e^{-\lambda}\lambda^i}{i!} = - \frac{e^{-\lambda}\lambda^{j}}{j!} = - \Pr(\Pois(\lambda) = j). \label{eq:deriv-1}\end{align}

From \eqref{eq:deriv}, it follows that
\begin{align}
\frac{d}{d\lambda} \bE[\min(\Pois(\lambda)&,k)](1+W_k/\lambda)\nonumber \\
& = \Pr(\Pois(\lambda) < k)(1 + W_k/\lambda) - \frac{W_k}{\lambda^2}\bE[\min(\Pois(\lambda),k)]\nonumber \\
& =\Pr(\Pois(\lambda) < k)(1 + W_k/\lambda) - \frac{W_k}{\lambda^2} \left(\lambda \Pr(\Pois(\lambda)<k)+k \Pr(\Pois(\lambda)>k) \right)\nonumber \\
& = \Pr(\Pois(\lambda) < k) - \frac{W_k}{\lambda^2}k\Pr(\Pois(\lambda)>k) \nonumber \\
& = \Pr(\Pois(\lambda)<k)\left(1-\frac{k^2}{\lambda^2}\frac{\Pr(\Pois(k)<k)}{\Pr(\Pois(k)>k)}\frac{\Pr(\Pois(\lambda)>k)}{\Pr(\Pois(\lambda)<k)}\right), \label{eq:derivativeTransformed}
\end{align}
The first equality follows from \eqref{eq:deriv}, the second from Lemma \ref{lem:poisson-facts}, the third from canceling terms, and the fourth from the definition of $W_k$ in \eqref{eq:wk}. 

 In~\eqref{eq:derivativeTransformed}, the derivative is clearly seen to be 0 when $\lambda=k$.
Our goal is to show that the derivative is positive when $\lambda<k$, and negative when $\lambda>k$. First suppose that $\lambda<k$.  Then
\begin{align*}
\frac{k^2}{\lambda^2}\frac{\Pr[\Pois(k)<k]}{\Pr[\Pois(k)>k]}\frac{\Pr[\Pois(\lambda)>k]}{\Pr[\Pois(\lambda)<k]}
&=\frac{\sum_{j>k}\lambda^{j-2}/j!}{\sum_{j>k}k^{j-2}/j!}\frac{\sum_{j<k}k^j/j!}{\sum_{j<k}\lambda^j/j!}
\\ &<(\frac{\lambda}{k})^{k-1}\frac{\sum_{j<k}k^j/j!}{\sum_{j<k}\lambda^j/j!}
\\ &=\frac{\sum_{j<k}\lambda^j(\frac{\lambda}{k})^{k-1-j}/j!}{\sum_{j<k}\lambda^j/j!}
\end{align*}
which is at most 1 because each term $(\frac{\lambda}{k})^{k-1-j}$ in the numerator is at most 1, for $j=1,\ldots,k-1$.
Substituting into~\eqref{eq:derivativeTransformed}, this shows that the derivative is positive when $\lambda<k$.

On the other hand, suppose that $\lambda>k$.  Then we can similarly derive
\begin{align*}
\frac{k^2}{\lambda^2}\frac{\Pr[\Pois(k)<k]}{\Pr[\Pois(k)>k]}\frac{\Pr[\Pois(\lambda)>k]}{\Pr[\Pois(\lambda)<k]}
&>(\frac{\lambda}{k})^{k-1}\frac{\sum_{j<k}k^j/j!}{\sum_{j<k}\lambda^j/j!}
\\ &=\frac{\sum_{j<k}\lambda^j(\frac{\lambda}{k})^{k-1-j}/j!}{\sum_{j<k}\lambda^j/j!}
\\ &\ge1.
\end{align*}
Substituting into~\eqref{eq:derivativeTransformed} shows that the derivative is negative when $\lambda>k$, completing the proof.
\end{myproof}

\section{Proofs for Theorem \ref{thm:a-m}}

\subsection{Proof of Lemma \ref{lem:lb}}

We begin by defining
\begin{equation} U(t,{\bf F}) =  \bE \limits_{\bV \sim \bF}\left[\sum_{i = 1}^n\max(V_i - t,0) \right] \end{equation}
Lemma \ref{lem:lb} follows immediately from Lemmas \ref{lem:lp} and \ref{lem:st-lb}.
\begin{lemma} \label{lem:lp}
For all $k \in \mathbb{N}$, $t \in \mathbb{R}_+$ and all $\bF$,
\[\OMN_k(\bF) \leq \LP_k(\bF) \leq t \cdot k + U(t,{\bf F}).\]
\end{lemma} 

\begin{myproof}[Proof of Lemma \ref{lem:lp}]
On any sample path, let $X_i\in\{0,1\}$ be the indicator for the prophet (who sees all value realizations up-front) accepting applicant $i$, for all $i$.
Note that we always have $\sum_i X_i\le k$.
Therefore, defining $x_i:=\bE[X_i]$ for all $i$ forms a feasible solution to the optimization problem defining $\LP_k(\bF)$.
Now, we have $\OMN_k(\bF)=\bE[\sum_iV_iX_i]=\sum_i x_i\bE[V_i|X_i=1]$ by definition.
However, $\bE[V_i|X_i=1]$ can be no greater than the average value of $V_i$ over its top $x_i$'th quantile, which can be written as $\frac{1}{x_i}\int_{1-x_i}^1 F^{-1}(z)dz$.
Therefore,
$$
\OMN_k(\bF)\le\sum_i x_i \frac{1}{x_i}\int_{1-x_i}^1 F^{-1}(z)dz=\sum_i\int_{1-x_i}^1 F^{-1}(z)dz
$$
which is at most $\LP_k(\bF)$ by the feasibility of $x_1,\ldots,x_n$.

To show the second inequality, note that for
any $t \in \mathbb{R}_+$, and any feasible solution $x_1,\ldots,x_n$ to the optimization problem defining $\LP_k(\bF)$ (i.e.\ satisfying $\sum_{i=1}^n x_i \le k;\ x_i\ge0\ \forall i$), its objective value satisfies
\begin{align}
\sum_{i=1}^n\int_{1-x_i}^1 F^{-1}(z)dz& =\sum_{i=1}^n\int_{1-x_i}^1 (F^{-1}(z)-t+t)dz \\
& \le\sum_{i=1}^n\int_{1-x_i}^1 \max(F^{-1}(z)-t,0)dz+\sum_{i=1}^n\int_{1-x_i}^1 t\cdot dz \\
& \le\sum_{i=1}^n\int_0^1 \max(F^{-1}(z)-t,0)dz+t\sum_{i=1}^n x_i \\
& \leq U(t,\bF)+ t k. \nonumber 
\end{align}
where we have applied the definition of $U(t,\bF)$ and the final inequality uses the fact that $t\ge0$ and $\sum_{i=1}^nx_i\le k$.
This completes the proof.
\end{myproof}

\begin{lemma} \label{lem:st-lb}
For all $k \in \mathbb{N}$, $t\in \mathbb{R}_+$ and all $\bF$, 
\[\ST_k^t(\bF) \geq t \cdot k \cdot \mathop{\bE}_{\bV \sim \bF}\left[\MR_k(D^t(\bV)) \right] + U(t,\bF) \mathop{\bE}_{\bV \sim \bF} \left[\AR_k(D^t(\bV)) \right].\]
\end{lemma}

In our proof, for fixed $t \in \mathbb{R}_+$ and $V \in \mathbb{R}_+^n$ we define 
\begin{align}
 D_i^t(V) & = \bI(V_i > t), \label{eq:Di}\\
 D^t(\bV) & = \sum_{i = 1}^n D_i(V),\label{eq:D}\\
D_{-i}^t(V) & = D^t(\bV) - D_i^t(V). \label{eq:D-i}
\end{align}
 The following identity can be quickly verified by considering the cases $D^t(\bV) < k$ and $D^t(\bV) \geq k$: 
\begin{equation} \min(D^t(\bV),k) = \sum_i D_i^t(V) \AR_k(D_{-i}^t(V)). \label{eq:tricky} \end{equation}
This identity is also used in the proof of Lemma \ref{lem:LPPerfBound}.

\begin{myproof}[Proof of Lemma \ref{lem:st-lb}]
Throughout, we fix $t$ and $V$, and write $D, D_i$ and $D_{-i}$, leaving the dependence on $t$ and $V$ implicit.

Suppose that $D_i = 1$. If competing demand $D_{-i}$ is less than $k$, then $i$ will certainly be accepted. Otherwise, $i$ is accepted with probability $k/(1+D_{-i})$. From this observation, we can lower bound the performance of any static threshold policy as follows.
\begin{align*}
\ST_k^t({\bf F})
& = \mathbb{E}\left[\sum_{i = 1}^n  V_i D_i \AR_k(D_{-i}) \right] \\
& = \mathbb{E} \left[\sum_{i = 1}^n  \left(tD_i + \max(V_i - t,0) \right) \AR_k(D_{-i}) \right] \\
& = t \cdot \mathbb{E}\left[\sum_{i = 1}^nD_i\AR_k(D_{-i})\right] + \mathbb{E}  \left[\sum_{i = 1}^n   \max(V_i - t,0) \AR_k(D_{-i}) \right] \\
& = t \cdot  \mathbb{E}\left[ \min(D,k)\right] + \sum_{i = 1}^n \mathbb{E}  \left[  \max(V_i - t,0) \right] \mathbb{E} \left[ \AR_k(D_{-i}) \right] \\
& \geq t \cdot  \mathbb{E}\left[ \min(D,k)\right] + \sum_{i = 1}^n \mathbb{E}  \left[  \max(V_i - t,0)  \right] \bE\left[ \AR_k(D)\right] \\
& = t \cdot k \cdot \bE\left[ \MR_k(D)\right] + U(t,{\bf F})  \bE\left[ \AR_k(D)\right].
\end{align*}
The fourth line uses \eqref{eq:tricky}, the fifth uses the fact that $D_{-i}$ and $V_i$ are independent, and the inequality uses the fact that $D \geq D_{-i}$ and $\AR_k$ is weakly decreasing.
\end{myproof}

\subsection{Proof of Lemma \ref{lem:structuralForMedian}} \label{sec:two-opt}

In this section we prove Lemma \ref{lem:structuralForMedian}.
Because the statement is trivially true if $n=1$, we assume $n\ge2$ and consider the problem of re-optimizing the pair of probabilities $p_1,p_2$ while holding all other probabilities fixed.

Let $D^-_{\vp}$ denote the sum of $n-2$ independent Bernoulli random variables with means $p_3,\ldots,p_n$. We note that for any function $f: \mathbb{N} \rightarrow \mathbb{R}$, 
\begin{align} 
\mathbb{E}[f(\Dp)]  &\!= \!(1\!-\!p_1)(1\!-\!p_2)\mathbb{E}[f(\Dp^-)]\!+\!(p_1(1\!-\!p_2)\!+\!p_2(1\!-\!p_1))\mathbb{E}[f(1\!+\!\Dp^-)]\!+\!p_1p_2\mathbb{E}[f(2\!+\!\Dp^-)] \nonumber
\\ &\!=\!\mathbb{E}[f(\Dp^-)] \!+\!(p_1\!+\!p_2)\mathbb{E}[f(1\!+\!\Dp^-)\!-\!f(\Dp^-)]\!+\!p_1 p_2 \mathbb{E}[f(2\!+\!\Dp^-)\!+\! f(\Dp^-)\!-\!2f(1\!+\!\Dp^-)].\label{eq:decomposition}
\end{align}

Note that the terms inside of the expectation operators do not depend on $p_1$ or $p_2$. This motivates the study of optimization problems of the following form.
\begin{align}
\min &\hspace{.3 in} B_0 + B_1 (p_1+p_2) + B_2 p_1 p_2 \label{eq:two-opt} \\
s.t. &\hspace{.3 in}  A_0 + A_1(p_1+p_2) + A_2 p_1p_2 = \phi \nonumber \\
&\hspace{.3 in}  (p_1, p_2) \in [0,1]^2. \nonumber
\end{align}

Lemma \ref{lem:two-opt} gives structural results about the solution to this class of optimization problem.
\begin{lemma} \label{lem:two-opt} 
If the optimization problem~\eqref{eq:two-opt} is feasible, then:
\begin{enumerate}
\item \label{part1} Any feasible solution in the interior $(0,1)^2$ with $p_1 \neq p_2$ is suboptimal, unless all feasible solutions have the same objective value. 
\item \label{part2}If $A_2 = 0, A_1 \neq 0$, and $B_2 < 0$, then any solution with $p_1 \neq p_2$ is suboptimal. 
\item \label{part3}If $A_2 \neq 0, A_1/A_2 \leq -1$, and $B_1 -  B_2 A_1/A_2 < 0$, then any solution with $p_1 \neq p_2$ is suboptimal.
\end{enumerate}
\end{lemma}
\begin{myproof}[Proof of \Cref{lem:two-opt}]
We prove the statements in \Cref{lem:two-opt} by analyzing several cases, depending on the values of the coefficients $A_i$ and $B_i$. 

Case 1 addresses cases where the constraint is not quadratic ($A_2 = 0$), and is relevant for the proof of Statements \ref{part1} and \ref{part2}. We divide this into two subcases: in Case 1a, the constraint is vacuous ($A_1 = 0$), while in 1b, it is linear. 

Case 2 addresses cases where the constraint is quadratic ($A_2 \neq 0$) and is relevant for the proof of Statements \ref{part1} and \ref{part3}. Case 2a corresponds to a case where at least one of two terms (after a variable transformation) is constrained to be 0.  Case 2b can be considered the ``general" case in which no critical expressions are equal to $0$.

{\bf Case 1a:} $A_2 = 0, A_1 = 0$. In this case, Statements \ref{part2} and \ref{part3} do not apply, so we prove only Statement \ref{part1}. If \eqref{eq:two-opt} is feasible, then its equality constraint is redundant. Note that the gradient of the objective function at a point $(p_1,p_2)$ is $(B_1+B_2p_2,B_1+B_2p_1)$. If $(p_1,p_2)$ is a local minimum, then this gradient must equal $(0,0)$. This can only occur if $p_1=p_2$ or $B_2=B_1=0$ (in which case all feasible points are optimal). This establishes Statement \ref{part1} in case 1a. 


{\bf Case 1b:} $A_2 = 0, A_1 \neq 0$. In this case, Statement \ref{part3} does not apply, so we prove only Statements \ref{part1} and \ref{part2}. In this case, we have $p_1+p_2 = (\phi-A_0)/A_1$, which makes $B_2p_1p_2$ the only variable term in the objective function to be minimized. If $B_2<0$, then the objective value is strictly reduced by setting $p_1$ and $p_2$ to be equal. This proves Statement \ref{part2} in case 1b. If $B_2 > 0$, then the objective value is strictly reduced by setting at least one of $p_1$ or $p_2$ to be an extreme point in $\{0,1\}$. If $B_2 = 0$, then all feasible solutions have the same objective value.
Combined, these arguments also establish Statement \ref{part1} in case 1b.

{\bf Case 2:} $A_2 \neq 0$. In this case, Statement \ref{part2} does not apply, so we prove only Statements \ref{part1} and \ref{part3}. Define 
\begin{align*}
q_i & = p_i + A_1/A_2 \text{ for } i \in \{1, 2\}; \\
\gamma & = \frac{\phi - A_0}{A_2} + \frac{A_1^2}{A_2^2}; \\
C_0 & = B_0 -2B_1A_1/A_2 +B_2(A_1/A_2)^2  + \gamma B_2; \\
C_1 & = B_1 - B_2A_1/A_2.
\end{align*}
Then~\eqref{eq:two-opt} can be reformulated as
\begin{align*}
\min & \hspace{.3 in} C_0 + C_1 (q_1 + q_2) \\
s.t. &  \hspace{.3 in} q_1q_2 = \gamma \\
& \hspace{.3 in} q_1, q_2 \in [A_1/A_2,A_1/A_2+1].
\end{align*}

{\bf Case 2a.} $A_2 \neq 0, \gamma = 0$. At least one of $q_1,q_2$ must be 0.  WLOG assume $q_1=0$ and consider the optimization problem over $q_2\in[A_1/A_2,A_1/A_2+1]$.  If $C_1<0$, then the unique optimal solution is $q_2=A_1/A_2+1$ (i.e. $p_2 = 1$); if $C_1>0$, then the unique optimal solution is $q_2=A_1/A_2$ (i.e. $p_2 = 0$); if $C_1=0$, then all feasible solutions are optimal. Moreover, if $C_1<0$ and $A_1/A_2 \le-1$, then $\gamma=0$ is only feasible if $A_1/A_2=-1$,  in which case the unique optimal solution is $q_1 = q_2 = 0$ (corresponding to $p_1 = p_2 = 1$). This establishes both Statements \ref{part1} and \ref{part3} in case 2a.

{\bf Case 2b.} $A_2 \neq 0, \gamma \neq 0$. In this case we cannot have $q_1 = 0$, so substitute $q_2=\gamma/q_1$, making the objective function $C_0+C_1(q_1+\gamma/q_1)$. If $C_1 = 0$, then all feasible solutions are optimal. If $C_1 \neq 0$, then the derivative of the objective can equal zero only if $1-\gamma/q_1^2=0$, in which case $q_1=q_2$ (and thus $p_1 = p_2$). Therefore, there is no local minimum with $(p_1, p_2) \in (0,1)^2$ and $p_1 \neq p_2$. It follows that when $C_1 \neq 0$, any global minimum either has $p_1 = p_2$ or lies on the boundary. This establishes Statement \ref{part1} in case 2b.

Turning to Statement \ref{part3}, note that if $A_1/A_2\le-1$, then $q_1, q_2 \leq 0$, so the problem is infeasible for $\gamma < 0$. Thus, assume $\gamma > 0$ and $q_1, q_2 < 0$. The second derivative of the objective equals $2C_1\gamma/q_1^3$ which is strictly positive. By strict convexity, the unique global minimizer must arise at the local minimum where $q_1=-\sqrt{\gamma}$. Furthermore, if $q_1 q_2 = \gamma$ has a solution in $[A_1/A_2,A_1/A_2 + 1]^2$, then it follows that $-\sqrt{\gamma} \in [A_1/A_2,A_1/A_2 + 1]$, and therefore the global minimizer is feasible. This establishes Statement \ref{part3} in case 2b.
\end{myproof}

In what follows, we use several established facts about the sum of independent Bernoulli random variables, which we collect in the following lemma. \na{Could state for $\Dp$ or $\Dp^-$. We use it for $\Dp^-$, but it seems most elegant (but perhaps more confusing) to state using $\Dp$.}
\begin{lemma} \label{lem:poisson-binomial}
Fix $n \in \mathbb{N}$ and $\vp \in [0,1]^n$, and let $\lambda = \mathbb{E}[\Dp] = \sum_{i = 1}^n p_i$. For $j \in \mathbb{N}_0$, let $h_j = \Pr(\Dp = j)$ and $H_j = \Pr(\Dp \le j)$. Then
\begin{enumerate}
\item \label{item:interval} The support  $\{j : h_j > 0\}$ is an interval.
\item \label{item:concave} The density $h$ is log-concave, meaning that for all $j \geq 1$,
\[h_j^2 \geq h_{j-1} h_{j+1}.\]
Furthermore, this inequality is strict whenever $h_j > 0$. 
\item \label{item:unimodal} The density $h$ is unimodal, with mode $\in \{\lfloor \lambda \rfloor, \lceil \lambda \rceil\}$. That is,
\begin{align*}
h_0 & \leq h_1 \leq h_2 \leq \cdots \leq h_{\lfloor \lambda \rfloor} \\
h_{\lceil \lambda \rceil} &  \geq  h_{\lceil \lambda \rceil+1} \geq h_{\lceil \lambda \rceil+2} \geq \cdots \\
\end{align*}
\item \label{item:cdf} For any $j \ge 2$,
\begin{equation} H_{j-2}h_{j-1} \leq H_{j-1} h_{j-2}.\end{equation}
\end{enumerate}
\end{lemma}

\begin{myproof}[Proof of \Cref{lem:poisson-binomial}]
The first fact is obvious. The second is shown by \citet{samuels_1965}, and the third is shown by several papers, including \citet{darroch_1964}, \citet{samuels_1965}, and \citet{wang_1993}.
The final fact follows because if the distribution of $D_{\vp}$ is log-concave, then so is the distribution of $n - D_{\vp}$, and log-concave distributions have increasing hazard rates (see \citet{an_1995}, Proposition 10).

\na{ \cite{samuels_1965} explicitly notes log-concavity. Unimodality has been proven by \cite{darroch_1964}, \cite{samuels_1965}, \cite{wang_1993}.\cite{jogdeo-samuels_1968} shows that the median is either $\lfloor \lambda \rfloor$ or $\lceil \lambda \rceil$. Many general facts about the Poisson Binomial distribution (sum of independent Bernoullis) are summarized by \cite{boland_2007} and \cite{tang-tang_2019}. \cite{baillon-cominetti-vaisman_2016} give a tight upper bound on its density function.}

\end{myproof}

\begin{myproof}[Proof of Lemma \ref{lem:structuralForMedian}]
If $\phi > 1$, then the optimization problem is infeasible. If $\phi = 1$, then the unique optimal solution is $p_i = 1$ for all $i$. Therefore, we assume that $\phi < 1$. It follows from the definition of $\MR$ in \eqref{eq:match-rate} that for any feasible solution, 
\begin{equation} \mathbb{P}(\Dp \leq k-1) > 0. \label{eq:d-minus-feas} \end{equation} 
Furthermore, we claim that for any optimal solution, we must have 
\begin{equation} \mathbb{E}[\AR_k(\Dp)] < 1. \label{eq:ar-opt} \end{equation}
This follows because $p_i = \phi/k$ for $i \in \{1, \ldots k\}$ and $p_i = 0$ for $i > k$ is a feasible solution (that is, $\mathbb{E}[\MR_k(\Dp)] = \phi$) and satisfies \eqref{eq:ar-opt}.

Take a feasible solution ${\bf p}$ in which $p_i \neq p_j$. We will show that this solution is not optimal. Without loss of generality, relabel so that $i = 1, j = 2$, and consider optimizing $p_1$ and $p_2$ with all other probabilities fixed. Let $h$ and $H$ denote the density and CDF of the sum of variables $\{3, 4, \ldots n\}$. We face an optimization problem in the form \eqref{eq:two-opt}, with
\begin{align}
A_1 &  = \mathbb{E}[\MR_k(1+D^-) - \MR_k(D^-)]  & \hspace{-.7 in} = H_{k-1}/k\,\,\, \label{eq:a1-new}\\
A_2 & = \mathbb{E}[\MR_k(2+D^-) +\MR_k(D^-)- 2\MR_k(1+D^-)]& \hspace{-.7 in}   = -h_{k-1}/k  \label{eq:a2-new}\\
B_1 &  = \mathbb{E}[\AR_k(1+D^-) - \AR_k(D^-)]  \label{eq:b1-new} \\
B_2 & = \mathbb{E}[\AR_k(2+D^-) +\AR_k(D^-)- 2\AR_k(1+D^-)]  \label{eq:b2-new}
\end{align}
Note that the second equalities for $A_1$ and $A_2$ above follow from the observations
\begin{align*}
\MR_k(1+j) - \MR_k(j)  & = \frac{1}{k}\bI(j \leq k-1),\\
\MR_k(2+j) + \MR_k(j) - 2 \MR_k(1+j) & =  -\frac{1}{k}\bI(j =k-1).
\end{align*}
Meanwhile, we have 
\begin{align}
\AR_k(1+j) - \AR_k(j)  & = \left\{ \begin{array}{c l} 0 & j \leq k-2 \\ \frac{-k}{(j+1)(j+2)} & j \geq k-1 \end{array} \right. \label{eq:delta1} \\
\AR_k(2+j) + \AR_k(j) - 2\AR_k(1+j) & = \left\{ \begin{array}{c l} 0 & j \leq k- 3 \\  \frac{-1}{k+1} & j = k-2 \\  \frac{2k}{(j+1)(j+2)(j+3)} & j \geq k-1 \end{array} \right. \label{eq:delta2}
\end{align}

We first consider the case where $A_2 = 0$, and then the case where $A_2 \neq 0$.

{\bf Case 1.} $A_2 = 0$. It follows from Lemma \ref{lem:poisson-binomial} part \ref{item:interval} that either 
\begin{equation} h_j = 0 \text{ for all } j \geq k -1, \label{eq:d-minus-small} \end{equation}
or $h_j = 0$ for all $j \leq k-1$. The latter case contradicts the feasibility constraint \eqref{eq:d-minus-feas}, so \eqref{eq:d-minus-small} must hold. It follows from \eqref{eq:d-minus-small} and \eqref{eq:a1-new} that $A_1 = 1/k$. Furthermore, \eqref{eq:b2-new} and \eqref{eq:delta2} imply $B_2 = -h_{k-2}/(k+1) \le 0$. If $B_2 = 0$, then $h_{k-2} = 0$, which further implies that $\Dp$ is at most $k-1$ and that $\mathbb{E}[\AR_k(\Dp)] = 1$. This contradicts the optimality condition \eqref{eq:ar-opt}. If $B_2 < 0$, then the second statement of Lemma \ref{lem:two-opt} implies that $\vp$ cannot be optimal.


{\bf Case 2.} $A_2 \neq 0$. By statement 3 of Lemma \ref{lem:two-opt}, $\vp$ is sub-optimal if the following conditions hold:
\begin{align}
A_1/A_2 \leq -1 \label{eq:a-ineq} \\ 
B_1 - B_2 A_1/A_2 < 0. \label{eq:b-ineq}
\end{align}
By \eqref{eq:a1-new} and \eqref{eq:a2-new}, $A_1/A_2 = -H_{k-1}/h_{k-1} \leq -1$, so \eqref{eq:a-ineq} holds.
All that remains is to show \eqref{eq:b-ineq}. For $j \in \mathbb{N}$, define
\[C_j = \frac{k(k+1)}{(j+1)(j+2)}, \]
and note that 
\[\frac{2k(k+1)}{(j+1)(j+2)(j+3)} = C_j - C_{j+1}. \]
By \eqref{eq:a1-new} and \eqref{eq:a2-new}, $A_2 \neq 0$ implies $H_{k-1} \geq h_{k-1} > 0$. By \eqref{eq:a1-new}, \eqref{eq:a2-new}, \eqref{eq:delta1} and \eqref{eq:delta2}, we have\begin{align} 
(k+1)(B_1 - B_2 A_1/A_2) & = - \sum_{j = k-1}^\infty C_j h_{j} + \frac{H_{k-1}}{h_{k-1}} \left(-h_{k-2} +  \sum_{j = k-1}^\infty (C_j - C_{j+1})h_{j}  \right) \nonumber \\
& = \frac{H_{k-2}}{h_{k-1}} \sum_{j = k-1}^\infty C_j h_{j} - \frac{H_{k-1}}{h_{k-1}}\left(h_{k-2} +  \sum_{j = k-1}^\infty C_{j+1}h_{j}\right)\nonumber \\
& = \frac{H_{k-1}}{h_{k-1}} \left( \frac{H_{k-2}}{H_{k-1}} \sum_{j = k-1}^\infty C_j h_{j} - \sum_{j =k-2}^\infty C_{j+1} h_j \right) \nonumber \\
& = \frac{H_{k-1}}{h_{k-1}} \sum_{j = k-1}^\infty \left( \frac{H_{k-2}}{H_{k-1}} h_{j}  -  h_{j-1} \right)C_j \label{eq:final-sum}
\end{align}
Part \ref{item:cdf} of Lemma \ref{lem:poisson-binomial} implies that
\[\frac{H_{k-2}}{H_{k-1}} \leq \frac{h_{k-2}}{h_{k-1}},\]
and therefore each term in \eqref{eq:final-sum} is non-positive, with some term strictly negative.
This implies that \eqref{eq:b-ineq} holds, and therefore by Statement 3 in Lemma \ref{lem:two-opt}, that ${\bf p}$ cannot be optimal. \na{Revisit this to explain why some term in \eqref{eq:final-sum} is strictly negative.}
\end{myproof}

\subsection{Proof of Lemma \ref{lem:pn-ub}}

\begin{myproof}[Proof of \Cref{lem:pn-ub}]
Applying \eqref{eq:favorite-id}, we see that for any demand distribution $D$,
\[\bE[\MR_k(D)] =\frac{1}{k} \bE[\min(D,k)]  = \frac{1}{k} \sum_{j = 0}^{k-1} \Pr(D > j)  = 1 - \frac{1}{k}\sum_{j = 0}^{k-1} \Pr(D \leq j).\]

Theorem 4 in \cite{hoeffding_1956} implies that for $j < k$, $\Pr(\Bin(n,k/n) \leq j)$ is decreasing in $n$, so 
\begin{align} \bE[\MR_k(\Bin(n,k/n)] & \geq \lim_{n \rightarrow \infty} \bE[\MR_k(\Bin(n,k/n)] \nonumber\\
& = \bE[\MR_k(\Pois(k))] \nonumber\\
& = \gamma_k \nonumber \\
& = \bE[\MR_k(\Bin(n,p_n)],\label{eq:chain} \end{align}
where the first equality follows from Le Cam's theorem, the second from  Lemma \ref{lem:poisson-facts}, and third from the definition of $p_{n,k}$ in \eqref{eq:pn-def}. Since $\MR_k$ is an increasing function, $\bE[\MR_k(\Bin(n,p)]$ is increasing in $p$, so \eqref{eq:chain} implies that $p_{n,k} \leq k$, completing the proof.
\end{myproof}

\section{Proofs for Theorems \ref{thm:a-lp} and \ref{thm:structuralGeneric}}

\subsection{Proof of Lemma \ref{lem:LPPerfBound}}

\begin{myproof}[Proof of Lemma \ref{lem:LPPerfBound}]
We fix a threshold $t \in \mathbb{R}_+$ and values $V \in \mathbb{R}_+^n$, and use $D$, $D_i$ and $D_{-i}$ as shorthand for $D^{t}(\bV)$, $D_i^{t}(V)$ and $D_{-i}^{t}(V)$ as defined in \eqref{eq:Di}, \eqref{eq:D} and \eqref{eq:D-i}. We claim that
\begin{align*}
k \cdot  \bE\left[ \MR_k(D)\right] & = \bE\left[ \min(D,k)\right]  \\
& = \mathbb{E}\left[ \sum_i D_i \AR_k(D_{-i}) \right] \\
& =  \sum_i \mathbb{E}[D_i] \mathbb{E}\left[\AR_k(D_{-i})\right] \\
& \geq  \sum_i \mathbb{E}[D_i]  \mathbb{E} \left[ \AR_k(D)\right] \\
& = \bE[D]  \bE\left[ \AR_k(D)\right].
\end{align*}
The second line uses \eqref{eq:tricky}, the third uses independence of $D_i$ and $D_{-i}$, the inequality uses that $D \ge D_{-i}$ and $\AR_k$ is decreasing, and the final line follows from \eqref{eq:D}, which states that $D = \sum_{i} D_i$. Therefore, if $t$ is chosen so that $\bE[D] = k$, it follows that $ \bE\left[ \MR_k(D)\right] \geq  \bE\left[ \AR_k(D)\right]$. 
\end{myproof}

\subsection{Proof of Theorem~\ref{thm:structuralGeneric} and Lemma \ref{lem:structuralForLP}}
\label{sec:structuralForLP}

\begin{myproof}[Proof of Theorem~\ref{thm:structuralGeneric}]
Suppose for contradiction that all optimal solutions have at least two distinct non-\{0,1\} probabilities. Select an optimal solution $\vp$ which maximizes the number of probabilities in \{0,1\}, which can be at most $n-2$ because there must exist indices $i,j$ for which $0<p_i<p_j<1$. Relabel these indices $i,j$ to be $1,2$, and consider the problem of optimizing $p_1$ and $p_2$, holding all others fixed. By \eqref{eq:decomposition}, this is a version of the optimization problem \eqref{eq:two-opt}. Because we have an interior optimal solution where the two probabilities are unequal, statement 1 of Lemma~\ref{lem:two-opt} implies that all feasible solutions must have the same objective value. Note that $p_1$ and $p_2$ can be modified to make one of them lie in $\{0,1\}$ while preserving feasibility. \na{Explain preceding claim?} This leads to an alternate optimal solution which has strictly more probabilities in \{0,1\}, causing a contradiction and completing the proof.
\end{myproof} 

\begin{myproof}[Proof of Lemma \ref{lem:structuralForLP}]
By Theorem \ref{thm:structuralGeneric}, there exists an optimal solution in which each $p_i \in \{0,p,1\}$, for some $p \in (0,1)$. Thus, to prove Lemma \ref{lem:structuralForLP}, it suffices to show that any feasible solution which has this property and has $p_i = p, p_j = 1$ for some $i,j$ is not optimal. 

Without loss of generality, relabel so that $i = 1$ and $j = 2$. Consider the problem of optimizing over $p_1$ and $p_2$, holding the remaining probabilities fixed. This is a version of the optimization problem \eqref{eq:two-opt}, in which $A_1 = 1, A_2 = 0$, and 
\[B_2 =  \mathbb{E}[\AR_k(2+D^-) + \AR_k(D^-)  - 2 \AR_k(1+D^-)]. \]
If $B_2 < 0$, then statement 2 in Lemma \ref{lem:two-opt} implies that our solution is not optimal. Therefore, all that remains is to show that $B_2 < 0$. Note that
\begin{equation} \AR_k(2+j) + \AR_k(j) - 2\AR_k(1+j) = \left\{ \begin{array}{c l} 0 & j \leq 3 \\  \frac{-1}{k+1} & j = k-2 \\  \frac{2k}{(j+1)(j+2)(j+3)} & j \geq k-1 \end{array} \right. \label{eq:delta2b} \end{equation}
For $j \geq k-1$ define 
\[C_j = \frac{2k(k+1)}{(j+1)(j+2)(j+3)} =  \frac{k(k+1)}{(j+1)(j+2)} - \frac{k(k+1)}{(j + 2)(j+3)}.   \]
Note that for $\ell \geq k-1$, we have
\begin{equation} \sum_{j = \ell}^\infty C_j =  \frac{k(k+1)}{(\ell+1)(\ell+2)}. \label{eq:c-equality} \end{equation}
In particular, taking $\ell = k-1$ we see that the $C_j$ sum to one. By \eqref{eq:delta2b} we have
\[ \mathbb{E}[\AR_k(2+D^-) + \AR_k(D^-)  - 2 \AR_k(1+D^-)]  = \frac{1}{k+1} \left( - h_{k-2} + \sum_{j = k-1}^\infty C_j h_{j}\right).\]
Showing that this is negative is equivalent to showing that $h_{k-2}$ is larger than a $C_j$-weighted average of $h_j$ for $j \geq k-1$.

Because $p_1 = p, p_2 = 1$, the constraint $\sum_{i = 1}^n p_i = k$ implies that $\mathbb{E}[D^-] = k-1-p$. Therefore, Lemma \ref{lem:poisson-binomial} part \ref{item:unimodal} implies that 
\begin{equation} h_j \leq h_k \text{ for all } j \geq k. \label{eq:ineq}\end{equation} 
From this and \eqref{eq:c-equality} we have
\begin{align}
- h_{k-2} + \sum_{j = k-1}^\infty C_j h_{j} & \leq - h_{k-2} +h_{k-1}C_{k-1} + h_{k}\sum_{j = k}^\infty C_j  \nonumber \\
& = -h_{k-2} + \frac{2}{k+2} h_{k-1} + \frac{k}{k+2} h_{k}. \label{eq:b2-neg}
\end{align}
We wish to establish that this expression is negative. Because indexes with $p_i = 0$ are irrelevant, we can assume without loss of generality that there are $n$ nonzero $p_i$. Let $d$ be the number of indices $i$ with $p_i = 1$ in our initial solution, so $n - d$ is the number of indices with $p_i = p$. Because $n-d \geq 1$ by assumption and $\sum_{i} p_i = d + p(n-d) = k$, we must have 
\begin{align}
d & \leq k-1, \label{eq:d-ineq} \\
p & = \frac{k-d}{n-d}. \label{eq:p-ineq}
\end{align}

Because $D^- - (d-1) \sim Bin(n-d-1,p)$, for $j \in \{d-1, d-2, \ldots, n-2\}$ we have
\begin{equation} h_{j} = {n-d-1 \choose j-d+1} p^{j-d+1} (1-p)^{n-j-2} \label{eq:dbinom}\end{equation}
Noting that $d-1 \leq k-2$ by \eqref{eq:d-ineq}, we can use \eqref{eq:dbinom} to rewrite the right side of \eqref{eq:b2-neg} as
$${n-d-1 \choose k-d} p^{k-d} (1-p)^{n-k-1}\left( -\frac{1-p}{p} \frac{k-d}{n-k}+ \frac{2}{k+2} + \frac{k}{k+2}\frac{p}{1-p} \frac{n-k-1}{k-d+1} \right).$$
The sign of this expression is determined by the second term. Substituting \eqref{eq:p-ineq} into this term, we see that it is equal to
\[-1+ \frac{2}{k+2}+ \frac{k}{k+2} \frac{n-k-1}{n-k} \frac{k-d}{k-d+1}.\]
This is clearly negative, as both fractions multiplying $\frac{k}{k+2}$ are strictly less than one.
\end{myproof}

\subsection{Proof of Lemma \ref{lem:annoying} for $k > 8$} \label{sec:pfsLP}

We begin with two technical lemmas.

\begin{lemma} \label{lem:varphiLB}
For any $k, n \in \mathbb{N}$ with $k \le n$, we have  $\bE[\AR_k(\Bin(n,\frac{k}{n}))]\ge1-\varphi_k(n)$, where
$$\varphi_k(n)=\left(1-\frac{k}{n+1}\right)\Pr[\Bin(n,\frac{k}{n})=k]+\frac{1}{2(n+1)}.$$
\end{lemma}
\begin{myproof}[Proof of \Cref{lem:varphiLB}]
We can derive that
\begin{align}
\bE[\AR_k(\Bin(n,k/n))]
&=\Pr[\Bin(n,k/n)<k]+\sum_{d=k}^n\frac{k}{d+1}\binom{n}{d}(k/n)^d(1-k/n)^{n-d} \nonumber
\\ &=\Pr[\Bin(n,k/n)<k]+\frac{n}{n+1}\sum_{d=k}^n\frac{(n+1)!}{(d+1)!(n-d)!}(k/n)^{d+1}(1-k/n)^{n-d} \nonumber
\\ &=\Pr[\Bin(n,k/n)<k]+\frac{n}{n+1}\Pr[\Bin(n+1,k/n)>k] \label{eqn:toPlot}
\\ &=\Pr[\Bin(n,k/n)<k]+\frac{n}{n+1}\left(\Pr[\Bin(n,k/n)>k]+\frac{k}{n}\Pr[\Bin(n,k/n)=k]\right) \nonumber
\\ &=1-\frac{1}{n+1}\Pr[\Bin(n,k/n)>k]-\left(1-\frac{k}{n+1}\right)\Pr[\Bin(n,k/n)=k] \nonumber
\end{align}
where the penultimate equality holds because by independence, $\Bin(n+1,k/n)>k$ if and only if either (i) there are at least $k$ successes in the first $n$ trials; or (ii) there are exactly $k$ successes in the first $n$ trials and the $n+1$'st trial is also successful.
The proof is then completed by the fact that the median of a $\Bin(n,k/n)$ random variable is $k$, which implies that $\Pr[\Bin(n,k/n)>k]\le1/2$.
\end{myproof}

\begin{lemma} \label{lem:derivLB}
For all integers $k\ge1$, treating $\varphi_k(n)$ as a function over real numbers in $n\in[k,\infty)$, we have that for all $n\ge k+2$,
\begin{align} \label{eqn:derivLB}
2(n+1)^2\varphi'_k(n)\ge k\frac{e^{-k}k^k}{k!}(1-\frac{1}{n}-\frac{1}{n-k}-\frac{1}{n(n-k)})-1,
\end{align}
where the RHS of~\eqref{eqn:derivLB} in increasing in $n$.
\end{lemma}
\begin{myproof}[Proof of \Cref{lem:derivLB}]
Define $\psi_k(n)=(1-\frac{k}{n+1})\Pr[\Bin(n,\frac{k}{n})=k]$, which equals the first term in $\varphi_k(n)$.
We can derive that
\begin{align*}
\psi_k(n)
&=\left(1-\frac{k}{n+1}\right)\frac{n(n-1)\cdots(n-k+1)}{k!}\frac{k^k}{n^k}\left(1-\frac{k}{n}\right)^{n-k}
\\ &=\frac{k^k}{k!}\left(1-\frac{k}{n+1}\right)\prod_{d=1}^{k-1}\left(1-\frac{d}{n}\right)\left(1-\frac{k}{n}\right)^{n-k}.
\end{align*}
Therefore,
\begin{align*}
\ln\psi_k(n) &=\ln\frac{k^k}{k!}+\ln(1-\frac{k}{n+1})+\sum_{d=1}^{k-1}\ln(1-\frac{d}{n})+(n-k)\ln(1-\frac{k}{n})
\\ &=\ln\frac{k^k}{k!}+\ln(n+1-k)-\ln(n+1)+\sum_{d=1}^{k-1}\ln(n-d)-(k-1)\ln n+(n-k)\ln(1-\frac{k}{n}).
\end{align*}

Taking the $\ln$ of $\psi_k$ helps us evaluate its derivative:
\begin{align*}
\frac{1}{\psi_k(n)}\psi'_k(n) &=\frac{1}{n+1-k}-\frac{1}{n+1}+\sum_{d=1}^{k-1}\frac{1}{n-d}-\frac{k-1}{n}+\ln(1-\frac{k}{n})+(n-k)\frac{1}{1-k/n}\frac{k}{n^2}
\\ &=\frac{1}{n+1-k}-\frac{1}{n+1}+\sum_{d=1}^{k-1}\frac{1}{n-d}-\frac{k-1}{n}+\ln(1-\frac{k}{n})+\frac{k}{n}
\\ &=\frac{1}{n+1-k}-\frac{1}{n+1}+\sum_{d=1}^{k-1}\frac{1}{n-d}+\frac{1}{n}-\ln\frac{n}{n-k}
\\ &=\frac{k}{(n+1)(n-k+1)}-\left(\int_{n-k}^n\frac{1}{x}dx-\sum_{d'=n-k+1}^n\frac{1}{d'}\right).
\end{align*}
Now, $\sum_{d'=n-k+1}^n1/d'$ represents a right Riemann sum for the integral $\int_{n-k}^n(1/x)dx$.
Since $1/x$ is decreasing, the area under the curve $1/x$ from $x=d'-1$ to $x=d'$ is greater than $1/d'$, for all $d'=n-k+1,\ldots,n$.
Moreover, since $1/x$ is convex, this difference is upper-bounded by the area of a triangle with base 1 and height $\frac{1}{d'-1}-\frac{1}{d'}$.
Therefore,
\begin{align*}
\frac{1}{\psi_k(n)}\psi'_k(n)
&\ge\frac{k}{(n+1)(n-k+1)}-\sum_{d'=n-k+1}^n\frac{1}{2}(\frac{1}{d'-1}-\frac{1}{d'})
\\ &=\frac{k}{(n+1)(n-k+1)}-\left(\frac{1}{2(n-k)}-\frac{1}{2n}\right)
\end{align*}
which implies that
\begin{align*}
\varphi'_k(n)
&\ge\psi_k(n)\left(\frac{k}{(n+1)(n-k+1)}-\frac{k}{2n(n-k)}\right)-\frac{1}{2(n+1)^2}
\\ &=k\Pr[\Bin(n,k/n)=k]\left(1-\frac{k}{n+1}\right)\left(\frac{2n(n-k)-(n+1)((n-k)+1)}{2n(n-k)(n+1)(n-k+1)}\right)-\frac{1}{2(n+1)^2},
\\ &=k\Pr[\Bin(n,k/n)=k]\left(\frac{n+1-k}{n+1}\right)\left(\frac{n(n-k)-n-(n-k)-1}{2n(n-k)(n+1)(n-k+1)}\right)-\frac{1}{2(n+1)^2},
\\ &=\frac{1}{2(n+1)^2}\left(k\Pr[\Bin(n,k/n)=k]\frac{n(n-k)-n-(n-k)-1}{n(n-k)}-1\right).
\end{align*}

We now use the fact that $\Pr[\Bin(n,k/n)=k]\ge e^{-k}k^k/k!$ to lower bound the expression in large parentheses by
\begin{align*}
k\frac{e^{-k}k^k}{k!}\left(1-\frac{1}{n}-\frac{1}{n-k}-\frac{1}{n(n-k)}\right)-1,
\end{align*}
noting that $1-\frac{1}{n}-\frac{1}{n-k}-\frac{1}{n(n-k)}\ge0$ when $n\ge k+2$.  Therefore,
\begin{align*}
\varphi'_k(n)\ge\frac{1}{2(n+1)^2}\left(k\frac{e^{-k}k^k}{k!}(1-\frac{1}{n}-\frac{1}{n-k}-\frac{1}{n(n-k)})-1\right),
\end{align*}
where the expression is large parentheses is now clearly increasing in $n$.
This completes the proof.
\end{myproof}

\begin{lemma} \label{lem:k31}
Lemma \ref{lem:annoying} holds for $k \ge 31$. 
\end{lemma}

\begin{myproof}[Proof of \Cref{lem:k31}]
Applying Lemma~\ref{lem:derivLB}, since the RHS of~\eqref{eqn:derivLB} is increasing in $n$, we see that for all $n\ge k+2$, we have
\begin{align*}
2(n+1)^2\varphi'_k(n)\ge k\frac{e^{-k}k^k}{k!}(1-\frac{1}{k+2}-\frac{1}{2}-\frac{1}{2(k+2)})-1,
\end{align*}
whose RHS can be numerically verified to be positive for all $k\ge31$.
Therefore, $\varphi_k(n)$ is increasing over $n\ge k+2$ and we have that
\begin{align} \label{eqn:0988}
\sup_{n\ge k+2}\varphi_k(n)=\lim_{n\to\infty}\varphi_k(n)=\frac{e^{-k}k^k}{k!}.
\end{align}

Meanwhile, by Lemma~\ref{lem:varphiLB}, we have that
\begin{align}
\inf_{n\ge k}\bE[\AR_k(\Bin(n,\frac{k}{n}))]
&\ge\min\{\bE[\AR_k(\Bin(k,\frac{k}{k}))],1-\varphi_k(k+1),\inf_{n\ge k+2}(1-\varphi_k(n))\} \nonumber
\\ &=\min\{\frac{k}{k+1},1-\frac{2}{k+2}\binom{k+1}{k}(\frac{k}{k+1})^k(\frac{1}{k+1})-\frac{1}{2(k+2)},1-\sup_{n\ge k+2}\varphi_k(n)\} \nonumber
\\ &=\min\{1-\frac{1}{k+1},1-\frac{2}{k+2}(1-\frac{1}{k+1})^k-\frac{1}{2(k+2)},1-\frac{e^{-k}k^k}{k!}\} \label{eqn:0628}
\end{align}
where the final equality applies~\eqref{eqn:0988}.
From~\eqref{eqn:0628}, the minimum can be numerically verified to equal the final argument for all $k\ge31$, where we note that the first two arguments are $1-O(1/k)$ while the final argument is $1-O(1/\sqrt{k})$.
\end{myproof}

\pagebreak

\begin{lemma} \label{lem:k8}
Lemma \ref{lem:annoying} holds for $9 \le k \le 30$. 
\end{lemma}

\begin{myproof}[Proof of \Cref{lem:k8}]
Applying Lemma~\ref{lem:derivLB}, since the RHS of~\eqref{eqn:derivLB} is increasing in $n$, we see that for all $n\ge k+11$, we have
\begin{align*}
2(n+1)^2\varphi'_k(n)\ge k\frac{e^{-k}k^k}{k!}(1-\frac{1}{k+11}-\frac{1}{11}-\frac{1}{11(k+11)})-1,
\end{align*}
whose RHS can be numerically verified to be positive for all $k\ge9$.
This demonstrates that $\varphi_k(n)$ is increasing over $n\ge k+11$.
However in fact, $\varphi_k(n)$ is increasing over integers $n\ge k+1$, for all $9\le k\le30$, as we numerically demonstrate in the table of values in Figure~\ref{fig:tableOfValues}.
Therefore, for all $9\le k\le30$, we have
\begin{align*}
\sup_{n\ge k+1}\varphi_k(n)=\lim_{n\to\infty}\varphi_k(n)=\frac{e^{-k}k^k}{k!},
\end{align*}
and applying Lemma~\ref{lem:varphiLB}, this implies that
\begin{align*}
\inf_{n\ge k}\bE[\AR_k(\Bin(n,\frac{k}{n}))]
&\ge\min\{\bE[\AR_k(\Bin(k,\frac{k}{k}))],\inf_{n\ge k+1}(1-\varphi_k(n))\}
\\ &=\min\{\frac{k}{k+1},1-\frac{e^{-k}k^k}{k!}\}.
\end{align*}
The minimum can be numerically verified to equal the final argument for all $9\le k\le30$.
\begin{figure}
\caption{Values of $\varphi_k(k+\ell)$, for all $9\le k\le30$ and $1\le\ell\le11$.  Note that the values in every row, which correspond to a fixed $k$, are increasing.}
\label{fig:tableOfValues}
\begin{tabular}{c|ccccccccccc}
$k$ & $\ell=1$ & $\ell=2$ & $\ell=3$ & $\ell=4$ & $\ell=5$ & $\ell=6$ & $\ell=7$ & $\ell=8$ & $\ell=9$ & $\ell=10$ & $\ell=11$ \\
\hline
9&0.1159&0.1164&0.1179&0.1193&0.1206&0.1216&0.1225&0.1233&0.1239&0.1245&0.1250\\
10&0.1059&0.1068&0.1086&0.1102&0.1116&0.1128&0.1138&0.1147&0.1154&0.1161&0.1167\\
11&0.0975&0.0987&0.1006&0.1024&0.1039&0.1052&0.1063&0.1073&0.1081&0.1088&0.1095\\
12&0.0904&0.0917&0.0938&0.0956&0.0973&0.0986&0.0998&0.1008&0.1017&0.1025&0.1032\\
13&0.0842&0.0857&0.0878&0.0897&0.0914&0.0928&0.0941&0.0951&0.0961&0.0969&0.0976\\
14&0.0788&0.0804&0.0826&0.0845&0.0862&0.0877&0.0890&0.0901&0.0910&0.0919&0.0927\\
15&0.0741&0.0758&0.0779&0.0799&0.0816&0.0831&0.0844&0.0855&0.0865&0.0874&0.0882\\
16&0.0699&0.0716&0.0738&0.0758&0.0775&0.0790&0.0803&0.0814&0.0825&0.0834&0.0842\\
17&0.0662&0.0679&0.0701&0.0720&0.0738&0.0753&0.0766&0.0777&0.0788&0.0797&0.0805\\
18&0.0628&0.0645&0.0667&0.0687&0.0704&0.0719&0.0732&0.0744&0.0754&0.0763&0.0772\\
19&0.0597&0.0615&0.0636&0.0656&0.0673&0.0688&0.0701&0.0713&0.0723&0.0733&0.0741\\
20&0.0570&0.0588&0.0609&0.0628&0.0645&0.0660&0.0673&0.0684&0.0695&0.0704&0.0713\\
21&0.0545&0.0562&0.0583&0.0602&0.0619&0.0633&0.0647&0.0658&0.0669&0.0678&0.0687\\
22&0.0522&0.0539&0.0560&0.0578&0.0595&0.0609&0.0622&0.0634&0.0644&0.0654&0.0662\\
23&0.0501&0.0518&0.0538&0.0556&0.0573&0.0587&0.0600&0.0612&0.0622&0.0631&0.0640\\
24&0.0481&0.0498&0.0518&0.0536&0.0552&0.0567&0.0579&0.0591&0.0601&0.0610&0.0619\\
25&0.0463&0.0480&0.0499&0.0517&0.0533&0.0547&0.0560&0.0571&0.0581&0.0591&0.0599\\
26&0.0446&0.0463&0.0482&0.0500&0.0515&0.0529&0.0542&0.0553&0.0563&0.0572&0.0581\\
27&0.0431&0.0447&0.0466&0.0483&0.0499&0.0512&0.0525&0.0536&0.0546&0.0555&0.0564\\
28&0.0416&0.0432&0.0451&0.0468&0.0483&0.0497&0.0509&0.0520&0.0530&0.0539&0.0547\\
29&0.0403&0.0419&0.0437&0.0453&0.0468&0.0482&0.0494&0.0505&0.0515&0.0524&0.0532\\
30&0.0390&0.0406&0.0423&0.0440&0.0455&0.0468&0.0480&0.0491&0.0500&0.0509&0.0517\\
\end{tabular}
\end{figure}

\end{myproof}

\section{Proofs of \Cref{prop:expDemandBad} and \Cref{prop:iid}} \label{sec:expDemand}

\begin{myproof}[Proof of \Cref{prop:expDemandBad}]
We prove this result in two cases. If $\phi < k$, then we can consider the degenerate case in which each variable is identically one. In this case, the static threshold policy accepts each item with probability $\phi/n$, and its expected value is the expected number of accepted items, which is $\bE[\min(k,\Bin(n,\phi/n))]$. As $n$ grows, the fact that $\phi < k$ implies that this value converges to a constant strictly less than $k \cdot \gamma_k$. Meanwhile, the prophet's value is equal to $k$. 

If $\phi > k$, then we can consider the case in which $n-1$ values are identically zero, and the remaining value is $n$ with probability $1/n$, and zero otherwise. Then the prophet's value is one. Meanwhile, when $n$ is large, the static threshold policy accepts each zero value item with probability approximately $\phi/n$, implying that demand is approximately Poisson distributed with mean $\phi$. It follows that when there is an item of value $n$, it is accepted with probability approximately $\bE[ \min(1,\frac{\phi}{1+\Pois(d)})]$, which is strictly less than $\gamma_k$ because $\phi > k \cdot \gamma_k$.
\end{myproof}

\begin{myproof}[Proof of Proposition \ref{prop:iid}]
Throughout, let $F$ denote the distribution from which each $V_i$ is drawn, and let $t$ denote a threshold such that expected demand is equal to $k$. This algorithm keeps $\min(D^t(\bV),k)$ values, each with expected value $\mathbb{E}_{\bV \sim F}[V \vert V > t]$. Meanwhile, by Lemma \ref{lem:lp}, 
$\OMN_k(\bF) \leq  tk + U(t,\bF) = k \cdot \mathbb{E}_{\bV \sim F}[V \vert V > t]$. Combining these, we see that
\[\frac{\ED_k(\bF)}{\OMN_k(\bF)} \geq \frac{\mathbb{E}[\min(D_{t}(V),k)] \cdot \mathbb{E}_{\bV \sim F}[V \vert V > t]}{k\cdot \mathbb{E}_{\bV \sim F}[V \vert V > t]}  = \frac{\mathbb{E}[\min(\Bin(n,k/n),k)]}{k}, \]
where we have used the fact that $D^t(\bV) \sim \Bin(n,k/n)$. By Le Cam's theorem, as $n \rightarrow \infty$, $\Bin(n,k/n) \rightarrow \Pois(k)$, and the ratio above converges to  
$\bE[\min(\Pois(k),k)]/k$, which equals $\gamma_k$ by Lemma \ref{lem:poisson-facts}.
\end{myproof}

\end{APPENDICES}
\end{document}